\documentclass[aps,prc,times,graphicx,tighten]{revtex4}
\usepackage[dvips]{graphicx}
\usepackage[dvips]{color}
\usepackage{comment}

\newcommand{\bea}{\begin{eqnarray}}
\newcommand{\eea}{\end{eqnarray}}
\newcommand{\be}{\begin{equation}}
\newcommand{\ee}{\end{equation}}
\newcommand{\np}{{\bf p}}

\newcommand{\nv}{{\bf v}}
\newcommand{\nuu}{{\bf u}}
\newcommand{\nh}{{\bf h}}
\newcommand{\nk}{{\bf k}}

\newcommand{\nq}{{\bf q}}

\begin{document}

\title{ 
Semi-inclusive two-nucleon emission in (anti) neutrino CC
  scattering within the relativistic mean field framework 
}

\author{
V.L. Martinez-Consentino$^a$,
A.M. Cantizani$^a$,
J.E. Amaro$^a$,
}

\affiliation{$^a$Departamento de F\'{\i}sica At\'omica, Molecular y Nuclear,
and Instituto de F\'{\i}sica Te\'orica y Computacional Carlos I,
Universidad de Granada, Granada 18071, Spain}

\date{\today}


\begin{abstract}
This paper delves into the distribution of semi-inclusive events
involving the emission of two nucleons in (anti) neutrino
charged-current scattering. The analysis is conducted within the
framework of relativistic mean field theory applied to nuclear
matter. To quantify the likelihood of such semi-inclusive events
occurring, we employ a relativistic model of meson-exchange currents
that aligns with the 2p2h inclusive cross-section. The outcomes are
presented in terms of one-fold and two-fold integrated semi-inclusive
cross sections. To highlight disparities among the various emission
channels, including proton-proton, neutron-proton, and 
neutron-neutron, we compare them against a purely phase-space
isotropic distribution within the center of mass of the two
nucleons. These comparisons reveal significant differences in the
event distributions, shedding light on the distinctive characteristics
of each channel.
 \end{abstract}


\maketitle

\section{Introduction}

The investigation of multinucleon emission reactions induced by
neutrinos and other electroweak probes has garnered significant
attention in contemporary physics
\cite{Gal11,Mor12,For12,Alv14,Mos16,Ath23}. This heightened interest stems
from the compelling evidence of its presence within the quasielastic
(QE) peak region, where a substantial contribution of two-nucleon
emission coexists with one-nucleon knockout events, defying
disentanglement in inclusive experiments. In other words, in this
context, the precise final hadronic state is largely unknown, except
for the possible absence of pions. This intriguing phenomenon has been
underscored in the analysis of neutrino and antineutrino scattering
experiments \cite{Agu10,Agu13,Fio13,Abe13}.

The significance of 2p2h processes in the inclusive cross section has
been underscored and substantiated through theoretical investigations
by various research groups \cite{Mar09,Nie11,Ama11,Lal12}.  These
studies employ diverse models that incorporate various nuclear
effects, such as meson-exchange currents (MEC) with $\Delta$-isobar
excitations, final-state interactions (FSI), short-range correlations
(SRC), the random-phase approximation (RPA) and effective
interactions.  These models encompass a range of approaches, including
Fermi gas models, which can be either local or global and may
incorporate relativistic corrections. Other approaches, on the other
hand, utilize shell models or quantum hadrodynamics models
\cite{Cuy16,Cuy17,Ber21}.  Ab initio methods also have unveiled that
MEC exert a substantial influence on the transverse response, a
phenomenon consistent with the significant presence of 2p2h
excitations. However, it's worth noting that in these calculations,
the contribution from distinct final states cannot be readily
disentangled \cite{Lov15,Lov20}.  In the factorization scheme, which
relies on the spectral function formalism, a similar contribution
attributed to 2p2h excitations has also been identified \cite{Roc19}.
However, the inclusion of these model-dependent ingredients has
resulted in noticeable disparities among the theoretical
predictions. As a consequence, numerous research endeavors have aimed
to intercompare the results to elucidate and mitigate the systematic
uncertainties inherent in neutrino data analyses
\cite{Gra13,Mar13b,Mar14,Ama12}.

This situation has necessitated the integration of two-nucleon
ejection mechanisms into Monte Carlo (MC) neutrino event generators
\cite{Cam22}. Typically, one commences with only the inclusive cross
section information in the 2p2h channel, as provided by theoretical
groups, and lacks the corresponding semi-inclusive cross section
data. In the absence of specific knowledge regarding the distribution
of the two final particles, it has become imperative to resort to
reasonable prescriptions for implementing an algorithm that generates
events with two-nucleon final states, based on given values of
momentum and energy transfer.
The conventional approach \cite{Sob12,GENIE,Kat15,Dol20} involves
selecting two nucleons from the Fermi sea. By ensuring energy-momentum
conservation, the four-momentum of the final hadronic state
(comprising two nucleons) can be computed. A reasonable assumption, in
the absence of more specific information, is to consider that the
distribution of the two final particles is isotropic in the center of
mass (CM) frame. In other words, within this frame, it is assumed that
the two final nucleons move in opposite directions with equal energy
and opposite momentum, and the emission angles are randomly chosen in the CM,
assuming an isotropic distribution. Once the final momenta
are determined, a boost is applied to transition to the laboratory
system, obtaining the momenta of the two ejected nucleons in this
frame. Subsequently, these momenta are further propagated within a
cascade FSI model \cite{Dyt21,Fra22}.

Therefore, while these event generators employ theoretical or
phenomenological methods to incorporate interactions during the
propagation of the final state, it is crucial to recognize that the
isotropic 2p distribution from which they begin is an assumption
that may not be entirely accurate and could introduce uncertainties
into the simulation results.
The distinction between the isotropic distribution employed in MC
generators and the more realistic distribution associated with a 2p2h
microscopic model is primarily delineated by the value of the
semi-inclusive hadronic tensor for a specific configuration of the two
final particles. This crucial ingredient is absent in most current
simulations. A recent endeavor to develop a microscopic model for
calculating the distribution of two outgoing nucleons, when compared to
the isotropic distribution model utilized in NEUT \cite{Sob20}, has revealed
significant discrepancies between the two distributions.
This discrepancy between the isotropic assumption and the actual
distribution of outgoing particles highlights a crucial area of
interest in the study of neutrino-induced two-nucleon emission
processes. It underscores the need for a more detailed and precise
understanding of the semi-inclusive hadronic tensor within such processes,
particularly with regard to the configurations of the final-state
nucleons.

The relevance of studying semi-inclusive reactions and implementing
them in Monte Carlo event generators is closely tied to the need for
improving the reconstruction of incident neutrino energy based on
measurements of the final state. This reconstruction would be possible
if both the final leptonic and hadronic states were known. Efforts in
this direction have included recent measurements of semi-inclusive
observables, which involve the kinematics of both the final lepton and
hadron(s), conducted by the T2K, MINERvA, and MicroBooNE
collaborations \cite{Abe18,Lu18,Cai20,Abr20,Abr20b}.

Simultaneously, a series of recent investigations have focused on the
theoretical \cite{Mor14,Van19,Fra20} study of semi-inclusive reactions
induced by neutrinos, where one nucleon is detected in coincidence
with the lepton \cite{Fra21,Bar21,Fra22,Fra22b,Fra23}. However, the
semi-inclusive one nucleon emission also has a contribution from 2p2h
processes and, in general, multinucleon emission. In the absence of a
semi-inclusive model for the 2p2h channel, attempts to estimate this
contribution have modeled it using the Monte Carlo implementation of
2p2h in GENIE \cite{Fra22b,Fra23}.  This, in turn, assumes that the
distribution of two nucleons is isotropic in the center of mass,
neglecting again the dependence of microscopical hadronic tensor on
the semi-inclusive variables. Therefore, this shows that there is a
need for theoretical studies of microscopic semi-inclusive versions
associated with the 2p2h cross sections.

In this paper, we embark on an in-depth exploration of this issue. Our
objective is to investigate the semi-inclusive cross section in
two-particle emission reactions induced by neutrinos, employing a
simple yet non-trivial microscopic model that consistently
incorporates the 2p2h hadronic tensor. This endeavor will enable us to
scrutinize the distribution of the two final nucleons and elucidate
how it deviates from the isotropic distribution in the CM
frame.  To achieve this, we will utilize a relativistic model of
electroweak MEC \cite{Pac03,Rui16}
 in conjunction with the Relativistic Mean Field (RMF) model of nuclear matter.
\cite{Ros80,Ser86,Weh93}. 

The MEC 2p2h responses, calculated within the framework of
the relativistic Fermi gas (RFG) model, have found extensive
application in numerous analyses and calculations of inclusive
neutrino and antineutrino cross sections, in conjunction with 
superscaling analysis (SuSA, SuSAv2) and
the spectral function model \cite{Meg16,Meg16b,Meg19,Iva18},
and have been implemented within the GENIE event generator as one of
the widely utilized 2p2h parametrization \cite{Dol20}.  

In the present work, we introduce several improvements to this 2p2h MEC
model. 
\begin{enumerate}
\item
First we use the RMF framework,
which accounts for the effects of the relativistic interaction between
nucleons via scalar and vector potentials, giving rise to a
relativistic effective mass and vector energy. 
\item
Second, we
include the complete $\Delta$ propagator, encompassing both its real
and imaginary parts.  The $\Delta$ propagator assumes a crucial role
within the MEC framework when a nucleon undergoes excitation to a
$\Delta$-isobar, subsequently decaying while interacting with a second
nucleon. This process 
constitutes a fundamental component of the
mechanisms underlying two-nucleon emission for high energy transfer.
\end{enumerate}

The 2p2h MEC model, based on the RMF framework,  considered in this work has been employed to compute
the inclusive 2p2h responses and cross sections in neutrino and
antineutrino charged-current (CC) scattering
\cite{Mar21,Mar21b,Mar23}.  This very model has also been recently
utilized to analyze the interference between the one-body current and
MEC contributions within the 1p1h response \cite{Cas23}. However, the
primary focus of this work lies in exploring the semi-inclusive
response associated with the emission of two particles, as deduced
from this microscopic model.  In the realm of semi-inclusive
reactions, the two emitted nucleons possess known momenta in the
proton-proton (PP), neutron-proton (NP), or neutron-neutron
channels. Yet, the state of the residual nucleus remains unknown. One
of the fundamental requirements for semi-inclusive distributions is
that their integral equals the inclusive 2p2h cross section. This
condition is essential to ensure the model's full consistency with the
inclusive framework, and it constitutes one of the fundamental tests
in this study.

The study of two-nucleon semi-inclusive reactions presents an
additional challenge due to the fact that the distributions of two
nucleons expand a six-dimensional space, making it impractical to
visualize results for the full two-nucleon distribution. A commonly
employed approach is to fix four or five of the final variables,
either angles or momenta, and then examine the distributions in the
remaining variables, typically one or two dimensions. However, this
method may not yield general conclusions unless the complete map of
configurations is explored, a task that is infeasible within the
confines of a few pages.
In this work, as it marks our initial exploration of these
observables, we present the results in an alternative manner. Our
approach involves fixing one or two final variables and integrating
over the remaining ones. This allows us to study partial
semi-inclusive cross sections, of the one-fold or two-fold type.
These one-fold and two-fold semi-inclusive cross sections encapsulate
the global information of the full 6D distribution, effectively
averaging over the remaining variables. This approach simplifies the
comparison between pure phase-space distributions and 
the model incorporating the hadronic tensor. Moreover,
it allows for more straightforward conclusions to be drawn, as any
differences observed in the one-fold or two-fold distributions imply
corresponding differences in the overall distributions.
By employing this strategy, we aim to provide a comprehensive view of
the semi-inclusive reactions induced by neutrinos, shedding light on
the intricacies of two-nucleon emission processes.

In Section II, we expound upon the formalism governing semi-inclusive
and inclusive two-nucleon knockout processes initiated by neutrinos.
In Section III, we delve into the method of calculation for both the
one-fold and two-fold cross sections employed in this study. This
method involves partial summations over bins within the space of
exclusive variables.  Section IV is devoted to the presentation of our
results, which encompass a range of lepton kinematics relevant to
neutrino scattering reactions.  Finally, in Section V, we draw our
conclusions, summarizing our findings and highlighting the key
insights gleaned from this study.

\section{Formalism}

\subsection{2p2h Cross section}

In this paper we consider the reactions $(\nu_\mu,\mu^- N_1 N_2)$ and
$(\overline{\nu}_\mu,\mu^+ N_1 N_2)$, where an incident (anti)
neutrino with energy $E_\nu$ interact with a complex nucleus changing
into a muon with kinetic energy $T_\mu$ and solid angle
$\Omega_\mu=(\theta_\mu,\phi_\mu)$, with the result of the ejection of
two nucleons, $N_1$, $N_2$, with momenta $(\np'_1,\np'_2)$ in the
final state plus a residual nucleus.  The outgoing nucleons can be
 neutrons or protons, $N_i=N,P$.  If the two nucleons are detected in
coincidence with the $\mu^\pm$ lepton the probability of the reaction
is determined by the semi-inclusive cross section
\begin{eqnarray}
\frac{d\sigma_{N_1N_2}}{d T_\mu d \Omega_\mu d^3p'_1 d^3p'_2}
=
\sigma_0(k,k')
L_{\mu \nu} 
W^{\mu \nu}_{\rm N_1 N_2}(\np'_1,\np'_2,\nq,\omega)
\end{eqnarray}
where $L_{\mu\nu}$ is the leptonic tensor and
$W^{\mu \nu}_{\rm N_1  N_2}(\np'_1,\np'_2,\nq,\omega)$
is the semi-inclusive hadronic tensor.
Denoting the neutrino and muon four-momenta as
$k^\nu=(\epsilon,\nk)$, and $k'{}^{\nu}=(\epsilon',\nk')$,
respectively, the momentum transfer is  $\nq=\nk-\nk'$
and the energy transfer is $\omega=\epsilon-\epsilon'$.
The four-momentum transfer $Q^\mu=(\omega,\nq$) verifies $Q^2=\omega^2-q^2<0$.

The function $\sigma_0$ in Eq. (1) is defined by
\begin{equation}
\sigma_0(k,k')=
\frac{G^2 \cos^2 \theta_c}{4 \pi^2} \frac{k'}{k},
\end{equation}
where $G=1.166\times 10^{-11}\quad\rm MeV^{-2}$ is the Fermi constant,
and $\theta_c$ is the Cabibbo angle, $\cos\theta_c=0.975$.  

The leptonic tensor  $L_{\mu\nu}$ appearing in Eq. (1) is 
given by 
\begin{eqnarray}\label{leptonic}
L_{\mu \nu}&=& 
 k_\mu k'_\nu + k_\nu k'_\mu-kk'g_{\mu\nu} \pm i
\epsilon_{\mu \nu \alpha \beta} k^\alpha k'^\beta,
\end{eqnarray}
where $+ (-)$ is for neutrino (antineutrino).

The semi-inclusive hadronic tensor 
$W^{\mu \nu}_{\rm N_1 N_2}(\np'_1,\np'_2,\nq,\omega)$
contains the information about the nuclear transition matrix
elements of the weak current in the above reaction.

The inclusive cross section in the two-nucleon emission 
channel is recovered by integration over the two final nucleons 
\begin{eqnarray}
\left(\frac{d\sigma}{d T_\mu d \Omega_\mu}\right)_{2p2h}
= 
\sigma_0
L_{\mu \nu}
\int  d^3p'_1 d^3p'_2
 W^{\mu \nu}(\np'_1,\np'_2,\nq,\omega) 
= 
\sigma_0
L_{\mu \nu}
 W^{\mu \nu}_{2p2h} (\nq,\omega),
\end{eqnarray}
with
\begin{equation}
  W^{\mu \nu} (\np'_1,\np'_2,\nq,\omega)
  = \sum_{N_1N_2}W^{\mu \nu}_{N_1N_2} (\np'_1,\np'_2,\nq,\omega),
\end{equation}
where the isospin sum is performed over final pairs (NP , PP)  in the case
of neutrino scattering, and (NN, NP) for antineutrinos.
The inclusive hadronic tensor in the two-particle, two-hole (2p2h) channel
is defined
\begin{equation}
 W^{\mu \nu}_{2p2h}(\nq,\omega) 
= \sum_{N_1N_2}\left(W^{\mu \nu}_{2p2h}\right)_{N_1N_2}(\nq,\omega)
\end{equation}
where the inclusive 2p2h hadronic tensor in the $(N_1,N_2)$ channel 
is 
\begin{equation}
\left(W^{\mu \nu}_{2p2h}\right)_{N_1N_2}(\nq,\omega)=
\int  d^3p'_1 d^3p'_2
 W^{\mu \nu} _{N_1N_2}(\np'_1,\np'_2,\nq,\omega).
\end{equation}

\begin{figure}
\includegraphics[width=8cm, bb=120 330 500 680]{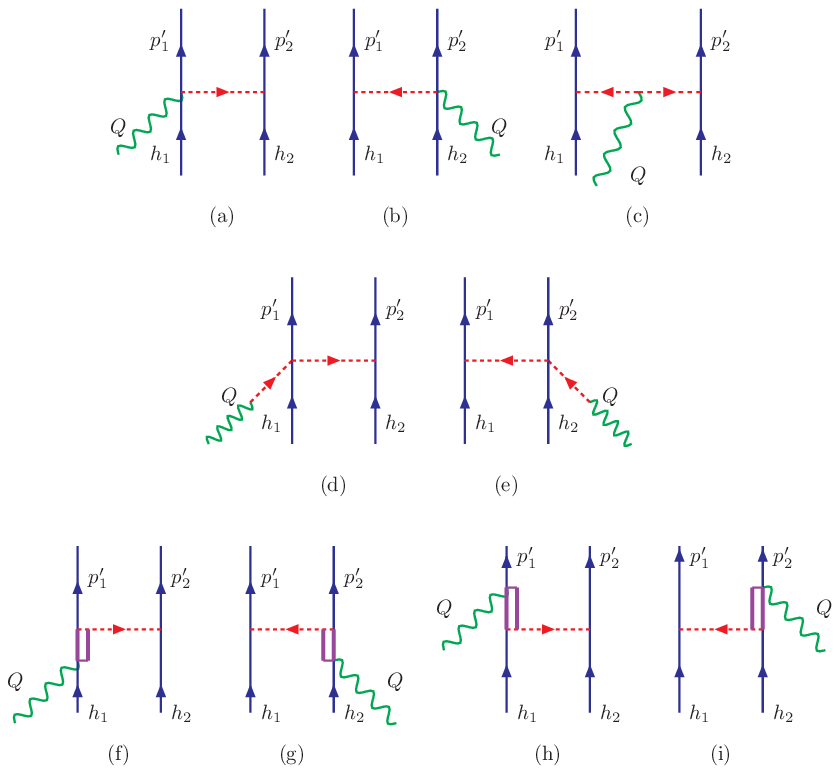}
\caption{Feynman diagrams of meson exchange currents 
considered in the present work.}
\label{diagmec}
\end{figure}

In Ref. \cite{Sim17} a MEC model was developed for
the inclusive 2p2h hadronic tensor in the 
RFG.  With independent-particle models such as the non-interacting
Fermi gas, the emission of two particles in a nuclear transition is
only possible through a two-body current operator. In this MEC model we
consider the diagrams with exchange of a pion as shown in Fig. 1.
This approach is based on the weak pion emission model of
\cite{Her07}.  Here we consider the same MEC model extended \cite{Mar21}
to include the
RMF within the framework of the
Walecka model \cite{Ros80,Ser86,Weh93}.  In this model the inclusive
2p2h hadronic tensor in the $N_1,N_2$ channel is written
 \begin{eqnarray}
(W^{\mu\nu}_{\rm 2p2h})_{N_1N_2}
&& 
=\frac{V}{(2\pi)^9}\int
d^3p'_1
d^3p'_2
d^3h_1
d^3h_2
\frac{(m^*_N)^4}{E_1E_2E'_1E'_2}
 w^{\mu\nu}_{N_1N_2}(\np'_1,\np'_2,\nh_1,\nh_2)
\delta(\np'_1+\np'_2-\nq-\nh_1-\nh_2)
\nonumber\\
&&
\times 
\theta(p'_1-k_F)\theta(k_F-h_1) 
\theta(p'_2-k_F)\theta(k_F-h_2) 
\delta(E'_1+E'_2-E_1-E_2-\omega),
\label{hadronic}
\end{eqnarray}
where 
$ w^{\mu\nu}_{N_1N_2}(\np'_1,\np'_2,\nh_1,\nh_2)$ is 
the elementary 2p2h hadronic tensor given below, Eq. (\ref{elementary}),
and  $V/(2\pi)^3 = Z/(\frac8 3 \pi k_F^3)$ for symmetric nuclear matter
 with Fermi momentum $k_F$.  
In Eq. (\ref{hadronic})
the variable  $m_N^*$ is the relativistic effective mass of the nucleon, 
the four-momenta of the final particles and holes are
$P'_i{}^\mu=(E'_i,\np'_i)$, and $H_i{}^\mu=(E_i,\nh_i)$, respectively.
Momentum conservation implies  $\bf p'_2= h_1+h_2+q-p'_1$.

Within the RMF model of nuclear matter
\cite{Ros80,Ser86,Weh93}, the nucleons are interacting with
relativistic scalar and vector potentials, denoted by $g_s\phi_0$, and
$g_vV_0$, see Ref. \cite{Ser86}.  The single particle wave functions
are plane waves with momentum $\np$, and with on-shell energy $E=
\sqrt{(m_N^*)^2+p^2}$.  The relativistic effective mass of the nucleon
is defined by $m_N^*=m_N-g_s\phi_0 = M^*m_N$, where $m_N$ is the free
nucleon mass.  Additionally the nucleon acquires a positive energy due
to the repulsion by the relativistic vector potential, $E_v=g_vV_0$.
Thus the total nucleon energy is $E_{RMF}=E+E_v$.  Note that the
vector energy does not appear explicitly in Eq. (\ref{hadronic}). This
is because it cancels out by subtraction of particle and hole energies
in the energy delta function. However it can be seen that the vector
energy enter in the $\Delta$ current diagrams (f--i) of Fig. 1,
through the energy of the intermediate $\Delta$ excitation  (see
Ref. \cite{Mar21} for details). 

The nuclear states in the RMF are Slater determinants constructed with
plane waves obtained by solving the free Dirac equation with effective
mass $m_N^*$.  Note that we use the same effective mass for particles
and holes.  All states with momentum $h<k_F$ are occupied in the
ground state.
The 2p2h excitations are
obtained by raising two particles above the Fermi level, with momenta
$p'_1$ and $p'_2>k_F$, leaving two holes with momenta $h_1$ and
$h_2<k_F$.  The 2p2h hadronic tensor is generated by the neutrino
interaction with the two-body MEC operator, whose matrix elements can
be written as
\begin{eqnarray}
\langle f | J^\mu(Q) | i \rangle =
\frac{(2\pi)^3}{V^2}\delta(\np'_1+\np'_2-\nq-\nh_1-\nh_2)
\frac{(m^*_N)^2}{\sqrt{E'_1E'_2E_1E_2}}
j^{\mu}(\np'_1,\np'_2,\nh_1,\nh_2).
\end{eqnarray}
The spin-isospin  two-body current function
$j^{\mu}(\np'_1,\np'_2,\nh_1,\nh_2)$ is
the sum of diagrams of Fig. 1, described in detail in
Refs. \cite{Sim17,Mar21}. 

The elementary 2p2h hadronic tensor
$w^{\mu\nu}_{N_1N_2}(\np'_1,\np'_2,\nh_1,\nh_2)$,
where $N_1,N_2$ are the
charge states of the final nucleons,
corresponds to the transition
\begin{equation}
  |\nh_1s_1,\nh_2s_2\rangle \rightarrow |\np'_1s'_1,\np'_2s'_2\rangle,
\end{equation}
where $s_i, s'_j$ are the initial and final spin components. 
We denote this transition, in short,
by $|1,2\rangle \rightarrow |1',2'\rangle$.
Then the elementary 2p2h hadronic tensor is given by 
\begin{equation}
  w^{\mu\nu}_{N_1N_2}(\np'_1,\np'_2,\nh_1,\nh_2)
  =
  \frac{1}{2}
\sum_{s_1s_2s'_1s'_2}
j^{\mu}(1',2',1,2)^*_A \,
j^{\nu}(1',2',1,2)_A \, , 
\label{elementary}
\end{equation}
where the two-body current
matrix element is anti-symmetrized with respect the pp or nn pair  
\begin{equation} \label{anti}
j^{\mu}(1',2',1,2)_A
\equiv
\left\{
\begin{array}{ccc}
j^\mu(1',2',1,2)- j^\mu(1',2',2,1),
\kern 1cm
&
\mbox{for $\nu_\mu NN \rightarrow \mu NP$} 
&
\mbox{or $\overline{\nu}_\mu PP \rightarrow \mu^+ NP$}, 
\\
j^{\mu}(1',2',1,2)- j^{\mu}(2',1',1,2),
\kern 1cm
&
\mbox{for $\nu_\mu NP \rightarrow \mu PP$} 
&
\mbox{or $\overline{\nu}_\mu NP \rightarrow \mu^+ NN$}. 
\\
\end{array}
\right.
\end{equation}
 The factor $1/2$ in Eq.~(\ref{elementary}) accounts for the
 anti-symmetry of the two-body wave function with respect
to  exchange of momenta and spin quantum numbers to avoid double counting.

\subsection{Probability distribution of 2p2h events}

In this work we are interested in studying the probability
distribution of semi-inclusive events corresponding to CC neutrino
(antineutrino) scattering, $E_\nu \rightarrow (T_\mu,\Omega_\mu)$, with
 two outgoing nucleons $N_1, N_2$ with momenta $\np'_1$,
$\np'_2$ in the final hadronic state.  In Monte Carlo event generators
the semi-inclusive event distribution is obtained by choosing two random
momenta for the holes, $\nh_1$, $\nh_2$, below $k_F$.  Then the total
momentum and energy of the final nucleons is computed as
$\np'=\nh_1+\nh_2+\nq$ and $E'=E_1+E_2+\omega$.  The individual momenta $\np'_1$
and $\np'_2$ for an event are generated by assuming an isotropic distribution of
the two final nucleons in the CM frame, and then they are transformed
to the Lab system. The 2p2h event 
${\cal E}=(\nh_1,\nh_2,\np'_1,\np'_2)$ is allowed if the conditions $p'_i>k_F$
are verified.  All 2p2h events ${\cal E}$ of this kind that are allowed for a
kinematics $(E_\nu,T_\mu,\Omega_\mu)$ contribute to the inclusive
cross section with some probability that depends on the value of the
hadronic tensor for the 2p2h event,
$w^{\mu\nu}(\np'_1,\np'_2,\nh_1,\nh_2)$.

To find such probability within our MEC model we proceed as follows.
We start with the inclusive 2p2h cross section written in the way
\begin{eqnarray}
 \frac{d\sigma_{N_1N_2} }{d T_\mu d \Omega_\mu}
   & =&
\frac{V}{(2\pi)^9}\int
d^3h_1
d^3h_2
\theta(k_F-h_1)\theta(k_F-h_2)
\frac{(m^*_N)^4}{E_1E_2}
F_{N_1N_2}(\nh_1,\nh_2),
\label{dsigma}
\end{eqnarray}
where the two-hole distribution function $F_{N_1N_2}(\nh_1,\nh_2)$ determines 
the contribution of each pair of holes $(\nh_1,\nh_2)$. It is given by 
\begin{eqnarray}
F_{N_1N_2}(\nh_1,\nh_2)=
\int
d^3p'_1
d^3p'_2
\frac{1}{E'_1E'_2}
\delta(E'_1+E'_2-E')
\delta(\np'_1+\np'_2-\np') 
f_{N_1N_2}(\np'_1,\np'_2,\nh_1,\nh_2),
\label{intlab}
\end{eqnarray}
where we have introduced the total energy and momentum in the final state,
$E'= E_1+E_2+\omega$ and  $\np'=  \nh_1+\nh_2+\nq$, respectively.
The function $f_{N_1N_2}(\np'_1,\np'_2,\nh_1,\nh_2)$ is
\begin{equation}
f_{N_1N_2}(\np'_1,\np'_2,\nh_1,\nh_2)=
\sigma_0 L_{\mu\nu}
w^{\mu\nu}_{N_1N_2}(\np'_1,\np'_2,\nh_1,\nh_2)
\theta(p'_1-k_F)\theta(p'_2-k_F).
\end{equation}
Note that
$P'{}^\mu=(E',\np')$
is the total four-momentum of the two-particle final state
$|\np'_1,\np'_2\rangle$ in the Lab frame. 
The integral over $(\np'_1,\np'_2)$
in Eq. (\ref{intlab}) can be performed by going to the center of mass 
system of the final nucleons. 
This was done in Ref. \cite{Rui14b} to compute the 2p2h phase space in 
frozen approximation (for $\nh_i=0$). 
Here we extend the method to arbitrary 
values of hole momenta and including the MEC dependence in the hadronic tensor. 

Doubly-primed variables refer to the CM system.
The total final momentum in the CM is zero.
$\np''= \np_1''+\np_2''=0$,
and the total final
energy $E''$ is determined by Lorentz invariance of the squared four-momentum,
$E''= \sqrt{E'^2-{p'}^2}$.

In the CM frame the two final nucleons are going
 back-to-back with the same momentum
and with the same energy
\begin{equation}
E''_1=E''_2= \frac{E''}{2} = \frac12 \sqrt{E'^2-{p'}^2} 
= \frac12\sqrt{(E_1+E_2+\omega)^2-(\nh_1+\nh_2+\nq)^2}.
\end{equation}
The condition $E''_1>m_N$ restricts the allowed 
values of $(\nh_1,\nh_2)$ 
for which two-nucleon emission is possible.

We perform the integration with respect CM coordinates, $\np''_1,\np''_2$,
using $d^3p'_i/E'_i= d^3p''_i/E''_i$.
Then we integrate over $\np''_2$ using the delta of momentum.
The two-hole distribution function, Eq. (\ref{intlab}), becomes 
\begin{eqnarray}
F_{N_1N_2}(\nh_1,\nh_2)=
\int
d^3p''_1
\frac{1}{(E''_1)^2}
\delta(2E''_1-E'')
f_{N_1N_2}(\np'_1,\np'_2,\nh_1,\nh_2) \theta(E'^2-p'^2-4m_N^{*2}),
\end{eqnarray}
where $\np''_2=-\np''_1$. 
The step function $\theta(E'^2-p'^2-4m_N^{*2})$
ensures that the invariant mass of the final two-particle hadronic system 
is larger than
$2m_N^*$ and also 
 that the total speed of
the CM is $v<1$.
Using $h_1''dh_1''=E_1''dE_1''$ to integrate
over the energy, $E''_1$ with the help of the Dirac delta function, we obtain
$E''_1=E''/2$, as expected. This fixes the value of the modulus
$p''_1=\sqrt{(E''_1)^2-(m_N^*)^2}$, resulting
\begin{eqnarray}
F_{N_1N_2}(\nh_1,\nh_2)=
\frac{p_1''}{2E_1''}
\int
d\Omega''_1
f_{N_1N_2}(\np'_1,\np'_2,\nh_1,\nh_2)\theta(E'^2-p'^2-4m_N^{*2}),
\label{intcm}
\end{eqnarray}
where $d\Omega''_1= d\cos\theta_1'' d\phi_1''$ are the angles of
$\np''_1$ in the CM system. Note that the function 
$f_{N_1N_2}(\np'_1,\np'_2,\nh_1,\nh_2)$ inside the integral (\ref{intcm}) 
has to be evaluated for the momenta
$\np'_1,\np'_2$ in the Lab system.
Then once $\np''_1,\np''_2$ are defined 
they are boosted back to the Lab system.

The CM moves with velocity $\nv=\np'/E'$. The direction of the velocity is defined by the unit vector $\nuu=\nv/v$. The boost of a CM vector $(E'',\np'')$ 
to the Lab system can be written using $\gamma= 1/\sqrt{1-v^2}$ 
\begin{eqnarray}
p'_u &=& \gamma(vE''+p''_u), \\
\np'_{\perp} &=& \np''_{\perp},
\end{eqnarray}
where $p'_u= \np'\cdot\nuu$ is the component along $\nuu$,
and $\np'_{\perp}$ is the invariant component perpendicular to $\nuu$.  
Using $\np'= p'_u\nuu + \np'_{\perp}$ we can write
\begin{equation}
\np'= \gamma(vE''+p''_u)\nuu +\np''_{\perp}
= (\gamma vE''+(\gamma-1)p''_u)\nuu +p''_u\nuu + \np''_{\perp}
= (\gamma vE''+(\gamma-1)p''_u)\nuu +\np''.
\end{equation}
Therefore the vectors $\np'_i$ in the Lab system 
inside the integral (\ref{intcm}) can be  computed as
\begin{equation}
\np'_i= 
 (\gamma vE''_i+(\gamma-1)\np''_i\cdot\nuu)\nuu +\np''_i.
\end{equation}
Inserting Eq. (\ref{intcm}) into (\ref{dsigma}) we can write the
2p2h inclusive cross section in the $N_1N_2$ channel as
\begin{eqnarray}
\left(\frac{d\sigma_{N_1N_2}}{d T_\mu d \Omega_\mu}\right)_{2p2h}
&    =&
\frac{V}{(2\pi)^9}\int
d^3h_1
d^3h_2
\theta(k_F-h_1)\theta(k_F-h_2)
\theta(E'^2-p'^2-4m_N^{*2})
\\
&&
\times \frac{(m^*_N)^4}{E_1E_2}
\frac{p_1''}{2E_1''}
\int
d\Omega''_1
f_{N_1N_2}(\np'_1,\np'_2,\nh_1,\nh_2).
\end{eqnarray}
  This integral over eight dimensions is
computed using numerical methods.  Note that with eight coordinates
$(h_1,\theta_1,\phi_1,h_2,\theta_2,\phi_2,\theta_r,\phi_r)$, we
generate the full space of exclusive events 
${\cal   E}=(\nh_1,\nh_2,\np'_1,\np'_2)$.  
For clarity here we denote by
$(\theta_r,\phi_r)$ the two nucleon angles relative to the CM, i.e.,
\begin{equation}
\theta_r=\theta''_1, \kern 1cm \phi_r=\phi''_1.
\end{equation}
Thus  the 2p2h inclusive cross section is finally written as 
\begin{eqnarray} \label{inclusiveD8}
\left( \frac{d\sigma_{N_1N_2}}{d T_\mu d \Omega_\mu}\right)_{2p2h}
   & =&
\int_{S_F}
d^3h_1
\int_{S_F}
d^3h_2
\int
d\cos\theta_r d\phi_r
G_{N_1N_2}(\nh_1,\nh_2,\theta_r,\phi_r),
 \label{inclusive}
\end{eqnarray}
where the integrals over holes is performed 
in the Fermi sphere, $S_F$, that is, for $h_i<k_F$, 
and the function $G_{N_1N_2}$ is
\begin{equation}  \label{gnn}
G_{N_1N_2}(\nh_1,\nh_2,\theta_r,\phi_r)
=\frac{V}{(2\pi)^9}
\frac{(m^*_N)^4}{E_1E_2}
\frac{p_1''}{2E_1''}
\sigma_0 L_{\mu\nu}
w^{\mu\nu}_{N_1N_2}(\np'_1,\np'_2,\nh_1,\nh_2)
\theta(p'_1-k_F)\theta(p'_2-k_F)
\theta(E'^2-p'^2-4m_N^{*2}),
\end{equation}
with 
\begin{equation}
E' = E_1+E_2+\omega, \kern 1cm p'=|\nh_1+\nh_2+\nq|,
\end{equation}
and the factor 
\begin{equation}
\frac{p_1''}{2E_1''}
= \frac12
\sqrt{
1-
\frac{4(m_N^*)^2}{(E_1+E_2+\omega)^2-(\nh_1+\nh_2+\nq)^2}
}
\end{equation}
comes from the Jacobian of the Lorentz transformation to the CM
system.  Note that the step function $\theta(E'^2-p'^2-4m_N^{*2})$
ensures that the value of above squared-root is real.  The function
$G_{N_1N_2}(\nh_1,\nh_2,\theta_r,\phi_r)$ determines the probability
distribution of exclusive events.

\subsection{Semi-inclusive 2p2h events}

Starting from equation (\ref{inclusiveD8}), the inclusive cross
section can be calculated numerically. We proceed by a discretization
procedure by selecting a representative set of the exclusive events
that contribute to the inclusive cross section.  With this set of
exclusive events all kinds of semi-inclusive events can be generated
and computed through partial sums. We must clarify the
difference between the concepts of exclusive event and semi-inclusive
event in the context of this approach. 

\begin{itemize}
\item By {\em exclusive event} we mean a set
of four vectors ${\cal E}=(\nh_1,\nh_2,\np'_1, \np'_2)$ 
that represent a particular excitation of
a 2p2h state that is compatible with conservation of energy and
momentum, and thus contributes to the integral (\ref{inclusiveD8}).
Each exclusive event, in turn, can be expressed as a set of 
two hole momenta and two relative angles
 in the center of mass system of the final nucleons
${\cal E}=(\nh_1,\nh_2,\theta_r, \phi_r)$. 
Note that each exclusive event carries an implicit, fixed value of the 
moment and energy
transfer $\nq,\omega$.

\item A {\em semi-inclusive event} refers to a fixed value of the
final momenta ${\cal E'}=(\np'_1,\np'_2)$ without specifying values
for the holes $(\nh_1 \nh_2)$, which are not observed. Therefore there
are many pairs of nucleons, $(\nh_1 \nh_2)$, that can contribute to a
given semi-inclusive event ${\cal E'}$, which implies in term of
probability or cross section a sum (or integral) over the contributing pairs,
 $(\nh_1 \nh_2)$, with the given restriction that they
belong to the semi-inclusive event ${\cal E'}$.

\item A {\em partial semi-inclusive event} is defined by specifying a
  subset of the six coordinates $(\np'_1,\np'_2)$ of the final state.
  In this work we will consider i) the {\em one-fold events} that
  correspond to fixing one of the values
  $(p'_1,\theta'_1,\phi'_1,p'_2,\theta'_2,\phi'_2)$.  And ii) the {\em
    two-fold events} that correspond to fixing two of them.  This will
  allow us to simplify the study of the semi-inclusive cross section,
  which depends on six variables, attacking first the simpler problem
  of the analysis of its partial integrals as will be seen below in
  more detail.

\item Finally, a {\em relative semi-inclusive} event refers to the
  specification of the two relative angles $(\theta_r, \phi_r)$ of the
  final particles, in the CM system of the final nucleons. It must be
  clarified that in a relative event of this type the total momentum
  of the final nucleons is not constant because many pairs of initial
  nucleons $(\nh_1, \nh_2)$ can contribute, and therefore the relative 
semi-inclusive events are
  not observable. But it will be useful to analyze the distribution of
  these relative events in order to study the influence of the
  hadronic tensor on the semi-inclusive cross section.

\end{itemize}

The probability distribution of partial semi-inclusive events 
that we will study in the next section
is determined by the following one-fold semi-inclusive cross sections
\begin{eqnarray}
\frac{d\sigma_{N_1N_2}}{d T_\mu d \Omega_\mu dp'_i},
\kern 1cm 
\frac{d\sigma_{N_1N_2}}{d T_\mu d \Omega_\mu d\cos\theta'_i},
\kern 1cm
\frac{d\sigma_{N_1N_2}}{d T_\mu d \Omega_\mu d\phi'_i},
\end{eqnarray}
for $i=1,2$, obtained by integration
 of the six-fold semi-inclusive cross section over five final variables. For instance 
\begin{eqnarray}
\frac{d\sigma_{N_1N_2}}{d T_\mu d \Omega_\mu dp'_1} 
&=&
\int 
      p'_1{}^2d\Omega'_1 d^3p'_2
\frac{d\sigma_{N_1N_2} }{d T_\mu d \Omega_\mu d^3p'_1d^3p'_2}.
\end{eqnarray}

The two-fold semi-inclusive cross sections are the following
\begin{eqnarray}
\frac{d\sigma_{N_1N_2}}{d T_\mu d \Omega_\mu dp'_1dp'_2},
&&
\frac{d\sigma_{N_1N_2}}{d T_\mu d \Omega_\mu d\cos\theta'_1d\cos\theta'_2},
\kern 1cm
\frac{d\sigma_{N_1N_2}}{d T_\mu d \Omega_\mu d\phi'_1d\phi'_2},
\\
\frac{d\sigma_{N_1N_2}}{d T_\mu d \Omega_\mu dp'_id\cos\theta'_j},
&&
\frac{d\sigma_{N_1N_2}}{d T_\mu d \Omega_\mu dp'_id\phi'_j},
\kern 1cm
\frac{d\sigma_{N_1N_2}}{d T_\mu d \Omega_\mu d\cos\theta'_id\phi'_j}.
\end{eqnarray}
with $i,j=1,2$.

We discretize the integration domain of Eq. (\ref{inclusiveD8}) to 
generate a finite set of bins of exclusive events.  
We choose the following integration
variables $(h_1^3,\cos\theta_1,\phi_1
,h_2^3,\cos\theta_2,\phi_2,\cos\theta_r,\phi_r)$.  We divide the
integration interval of each variable into a finite number of
sub-intervals of respective widths $(\Delta
h^3,\Delta\cos\theta,\Delta\phi,\Delta
h^3,\Delta\cos\theta,\Delta\phi,\Delta\cos\theta_r,\Delta\phi_r)$.
Note that we use the variable $h^3$ instead of $h$ to perform the
discretization, 
that goes from $0$ to $k_F^3$.  This is convenient
because the property $h^2dh = dh^3/3$ to improve the precision in the 
numerical integral. We use the same integration widths for the two initial
nucleons.
In this way
the space of exclusive events is discretized as a finite set $\{ {\cal E}_i |
i=1,\ldots {\cal N}\}$.  The volume element in the dicretized
space of exclusive events is
\begin{equation}
\Delta {\cal E} = 
\left( \frac 13 \Delta h^3 \Delta\cos\theta_h\Delta\phi_h  \right)^2
\Delta\cos\theta_r\Delta\phi_r.
\end{equation}
The inclusive cross section can be approximated as a sum over 
discrete exclusive events 
\begin{equation} \label{discreta}
  \left(\frac{d\sigma_{N_1N_2}}{d T_\mu d \Omega_\mu}\right)_{2p2h}
 \simeq 
\sum_{i=1}^{\cal N} G_{N_1N_2}({\cal E}_i) \Delta {\cal E}. 
\end{equation}
In the limit $\Delta {\cal E} \rightarrow 0$ the 
exact cross section is obtained.

According to equation (\ref{discreta}) 
 the probability  of each exclusive event,
for a given lepton kinematics $E_\nu,T_\mu,\Omega_\mu$  is given by
\begin{equation} \label{probE}
P_{N_1N_2}({\cal E}) = \frac
 { G_{N_1N_2}({\cal E}) \Delta {\cal E}} 
 {\displaystyle
 \left(\frac{d\sigma_{N_1N_2}}{d T_\mu d \Omega_\mu}\right)_{2p2h} },
\end{equation}
and it is normalized to one
\begin{equation} \label{normaliza}
\sum_{i=1}^{\cal N} P_{N_1N_2}({\cal E}_i) =1.
\end{equation}
The function $G_{N_1N_2}({\cal E})$ is given by Eq. (\ref{gnn}) were
we see that for fixed values $\nh_1$ and $\nh_2$ the outgoing nucleons
are distributed according to the available phase space but they are
{\em not} generated uniformly in the center of mass of the final
hadronic system, because the function $G_{N_1N_2}({\cal E})$ depends
on the value of the exclusive 2p2h hadronic tensor, $w^{\mu\nu}({\cal
  E})$ for the event ${\cal E}$.  This contrasts with the procedure
used in most model implementations in neutrino interaction event
generators, where an isotropic distribution for the two outgoing
nucleons is assumed.  This is the case for instance of the NuWro
\cite{Sob12}, NEUT \cite{Sob20}, and GENIE \cite{Dol20} event
generators. Something similar happens in GiBUU implementation of the
2p2h excitations where the exclusive hadronic tensor does not depend
on the event ${\cal E}$ \cite{Lal12}.  In the results section
we will study  the effect of including or not the exclusive
hadronic tensor in the distribution of the final particles and the
difference between the distributions in the different charge channels
(PN, PP or NN).

\section{One- and two-fold 2p2h semi-inclusive cross sections}

To calculate the one- and two-fold semi-inclusive cross sections, we
begin by generating the discrete set of exclusive events corresponding
to a specific lepton kinematics.  This means a uniform set of
coordinates
$(h_1^3,\cos\theta_1,\phi_1,h_2^3,\cos\theta_2,\phi_2,\cos\theta_r,\phi_r)$
for holes and relative angles of the particles, providing a discreet
set of exclusive events ${\cal E}_i$ that generates the inclusive
cross section. In fact, the first check we make is that the sum
over events of the function $G({\cal E})$, Eq. (\ref{discreta}), reproduces the
inclusive 2p2h cross section for the given kinematics.

Discretization in the space of exclusive events 
implies in particular the discretization of semi-inclusive 
events  $(\np'_1,\np'_2)$. 
We divide the intervals of possible values 
of the exclusive variables into $n$ sub-intervals or 
{\em bins}:
\begin{eqnarray}
p'_i &: & [ k_F, (p'_i)_{\rm max} ] = 
[ p'_i{}^{(1)}, p'_i{}^{(2)}, \ldots, p'_i{}^{(n+1)} ],    
\\
\cos\theta'_i  &:&  [-1,1] = 
[ \cos\theta'_i{}^{(1)}, \cos\theta'_i{}^{(2)},\ldots,\cos\theta'_i{}^{(n+1)}],
\\
\phi'_i  &:&  [0, 2 \pi ] = 
[ \phi'_i{}^{(1)}, \phi'_i{}^{(2)}, \ldots, \phi'_i{}^{(n+1)}],    
\end{eqnarray} 
where the maximum momentum of final nucleons is
 $(p'_i)_{\rm max}=\sqrt{(E'_i)^2_{\rm max}-(m_N^*)^2}$ with 
$(E'_i)_{\rm max}=E_F+\omega$, corresponding to a nucleon with Fermi energy 
that receives all the energy transfer.   

\subsection{One-fold semi-inclusive cross sections}

The one-fold semi-inclusive cross sections can be computed for each bin
as follows. 

Let $X$ be one of the semi-inclusive variables $X= p'_i,\cos\theta'_i,\phi'_i$,
for $i=1,2$.
For each exclusive event ${\cal E}$
 we will denote by $X({\cal E})$  the corresponding coordinate of the event. 
For instance if ${\cal E}=(\nh_1,\nh_2,\np'_1,\np'_2)$ is an event, then 
$p'_1({\cal E})= p'_1$.
Let us now define by 
$B(X^{(k)})$ 
the subset of events ${\cal E}$ such as 
$X^{(k)}\leq X({\cal E}) < X^{(k+1)}$.
That is, the event ${\cal E}$ belong to the $k$-th bin of the variable $X$, 
i.e, the interval $[X^{(k)},X^{(k+1)}]$ 
\begin{equation}
B(X^{(k)}) \equiv 
\left\{ {\cal E} \kern 5mm \mbox{such that} \kern 5mm
X^{(k)}\leq X({\cal E}) < X^{(k+1)}
\right\}.
\end{equation}
The total probability that an event belong to the bin $k$ 
is obtained by summing the probabilities 
of all events that fall within the bin
\begin{equation}
P_{N_1N_2}(X^{(k)})= 
\sum_{{\cal E}\in B(X^{(k)})}P_{N_1N_2}({\cal E}), 
\end{equation}
verifying
\begin{equation} \label{prob1}
\sum_k P_{N_1N_2}(X^{(k)})=  
\sum_k \sum_{{\cal E}\in B(X^{(k)})}P_{N_1N_2}({\cal E}) =1. 
\end{equation}
Let be  $\Delta X= X^{(k+1)}-X^{(k)}$ the (constant) width of the bin
of the $X$ variable. We define
\begin{equation} \label{probability}
S(X^{(k)})=
 \frac{P_{N_1N_2}(X^{(k)})}{\Delta X} 
\left(\frac{d\sigma_{N_1N_2}}{d T_\mu d \Omega_\mu}\right)_{2p2h}.
\end{equation}
Using Eq (\ref{prob1}) we obtain
\begin{equation}
\sum_k S(X^{(k)})\Delta X 
=
\left(\frac{d\sigma_{N_1N_2}}{d T_\mu d \Omega_\mu}\right)_{2p2h}.   
\end{equation}
Therefore we identify  $S(X^{(k)})$ as the corresponding
 one-fold semi-inclusive cross section averaged over the bin $X^{(k)}$.
\begin{equation} \label{one-fold}
\left.\frac{d\sigma_{N_1N_2}}{d T_\mu d \Omega_\mu dX}\right|_{X^{(k)}}   
= S(X^{(k)}) = 
\frac{1}{\Delta X}
\sum_{{\cal E}\in B(X^{(k)})} G_{N_1N_2}({\cal E}) \Delta {\cal E}. 
\end{equation}
By summation over semi-inclusive bins 
we recover the inclusive 2p2h cross section  
\begin{eqnarray}
\sum_k 
\left.\frac{d\sigma_{N_1N_2}}{d T_\mu d \Omega_\mu dX}\right|_{X^{(k)}}   
\Delta X
& = &
\frac{d\sigma_{N_1N_2}}{d T_\mu d \Omega_\mu }.
\end{eqnarray}
By this approach we compute the six one-fold 2p2h semi-inclusive cross sections
as
\begin{eqnarray}
\left.\frac{d\sigma_{N_1N_2}}{d T_\mu d \Omega_\mu dp'_i}\right|_{p'^{(k)}_i}   
&=& 
\frac{1}{\Delta p'_i}
\sum_{{\cal E}\in B(p'^{(k)}_i)} G_{N_1N_2}({\cal E}) \Delta {\cal E}, 
\\
\left.
\frac{d\sigma_{N_1N_2}}{d T_\mu d \Omega_\mu d\cos\theta'_i}
\right|_{\cos\theta'^{(k)}_i}   
&=& 
\frac{1}{\Delta \cos\theta'_i}
\sum_{{\cal E}\in B(\cos\theta'^{(k)}_i)} G_{N_1N_2}({\cal E}) \Delta {\cal E}, 
\\
\left.\frac{d\sigma_{N_1N_2}}{d T_\mu d \Omega_\mu d\phi'_i}\right|_{\phi'^{(k)}_i}   
&=& 
\frac{1}{\Delta \phi'_i}
\sum_{{\cal E}\in B(\phi'^{(k)}_i)} G_{N_1N_2}({\cal E}) \Delta {\cal E}, 
\\
\end{eqnarray}
where the partial sums are performed over the subset of  events
corresponding to the bins of $p'_i,\cos\theta'_i$ or $\phi'_i$:
\begin{eqnarray}
B(p'_i{}^{(k)}) 
&=&
\left\{ {\cal E} \kern 5mm \mbox{such that} \kern 5mm
p'_1{}^{(k)}\leq p'_1({\cal E}) < p'_1{}^{(k+1)}
\right\},
\\
B(\cos\theta'_i{}^{(k)}) 
&=&
\left\{ {\cal E} \kern 5mm \mbox{such that} \kern 5mm
\cos\theta'_1{}^{(k)}\leq \cos\theta'_1({\cal E}) < \cos\theta'_1{}^{(k+1)}
\right\},
\\
B(\phi'_i{}^{(k)}) 
&=&
\left\{ {\cal E} \kern 5mm \mbox{such that} \kern 5mm
\phi'_1{}^{(k)}\leq \phi'_1({\cal E}) < \phi'_1{}^{(k+1)}
\right\}.
\\
\end{eqnarray}

\subsection{Two-fold semi-inclusive cross sections}

The above procedure to obtain the one-fold semi-inclusive cross sections
is straightforwardly extended to the case of 
two-fold semi-inclusive cross sections.
If $X \ne Y$ is a pair of 
semi-inclusive variables $X,Y= p'_i,\cos\theta'_i,\phi'_i$,
for $i=1,2$, we construct
the subset
$B(X^{(k)},Y^{(l)})$ 
of exclusive events ${\cal E}$ such that 
$X({\cal E})$ is inside the $k$-th bin of the variable $X$, 
and $Y({\cal E})$ is inside the $l$-th bin of the variable $Y$, 
i.e, 
\begin{equation}
B(X^{(k)}, Y^{(l)}) 
\equiv 
\left\{ {\cal E} \kern 5mm |  \kern 5mm
X^{(k)}\leq X({\cal E}) < X^{(k+1)} \, \mbox{and}\,
Y^{(l)}\leq Y({\cal E}) < Y^{(l+1)}
\right\}.
\end{equation}
The total probability that an event belong to  the bins $(k,l)$
of variables $(X,Y)$ is then  
\begin{equation}
P_{N_1N_2}(X^{(k)}, Y^{(l)})= 
\sum_{{\cal E}\in B(X^{(k)},Y^{(l)})}P_{N_1N_2}({\cal E}). 
\end{equation}
Again the total probability is one
\begin{equation} \label{prob2}
\sum_{kl} P_{N_1N_2}(X^{(k)},Y^{(l)})=  
\sum_{kl} \sum_{{\cal E}\in B(X^{(k)},Y^{(l)})}P_{N_1N_2}({\cal E}) =1. 
\end{equation}
As in the previous subsection, we define
\begin{equation} \label{probability2}
S(X^{(k)},Y^{(l)})=
 \frac{P_{N_1N_2}(X^{(k)},Y^{(l)} ) }{\Delta X \Delta Y} 
\left(\frac{d\sigma_{N_1N_2}}{d T_\mu d \Omega_\mu}\right)_{2p2h}.
\end{equation}
Using Eq (\ref{prob2}) we obtain
\begin{equation}
\sum_{kl} S(X^{(k)},Y^{(l)}) \Delta X  \Delta Y 
=
\left(\frac{d\sigma_{N_1N_2}}{d T_\mu d \Omega_\mu}\right)_{2p2h}.   
\end{equation}
Therefore we identify  $S(X^{(k)},Y^{(l)})$ as the corresponding
two-fold semi-inclusive cross section averaged over the 
the  bins $X^{(k)},Y^{(l)}$ 
\begin{equation}\label{two-fold}
\left.\frac{d\sigma_{N_1N_2}}{d T_\mu d \Omega_\mu dX dY}\right|_{X^{(k)},Y^{(l)}}   
= S(X^{(k)},Y^{(l)}) =
\frac{1}{\Delta X\Delta Y}
\sum_{{\cal E}\in B(X^{(k)},Y^{(l)})} G_{N_1N_2}({\cal E}) \Delta {\cal E}. 
\end{equation}
Is easy to check that the  summation over two-fold semi-inclusive bins,
 gives again the inclusive 2p2h cross section  
\begin{eqnarray}
\sum_{kl} 
\left.
\frac{d\sigma_{N_1N_2}}{d T_\mu d \Omega_\mu dX dY}
\right|_{X^{(k)},Y^{(l)}}   
\Delta X \Delta Y
& = &
\frac{d\sigma_{N_1N_2}}{d T_\mu d \Omega_\mu }.
\end{eqnarray}
Therefore the probability distribution of the two-fold events is related to the two-fold semi-inclusive cross section by:
\begin{equation} \label{probability3}
\left.
\frac{d\sigma_{N_1N_2}}{ d T_\mu d \Omega_\mu dX dY}
\right|_{X^{(k)},Y^{(l)} }   =
 \frac{P_{N_1N_2}(X^{(k)},Y^{(l)})}{\Delta X \Delta Y} 
\left(\frac{d\sigma_{N_1N_2}}{d T_\mu d \Omega_\mu}\right)_{2p2h}.
\end{equation}

Using Eq. (\ref{two-fold}), the
 two-fold 2p2h semi-inclusive cross sections are computed as 
\begin{eqnarray}
\left.
\frac{d\sigma_{N_1N_2}}{d T_\mu d \Omega_\mu dp'_1 dp'_2}
\right|_{p'^{(k)}_i ,  p'^{(l)}_j}   
&=& 
\frac{1}{\Delta p'_1\Delta p'_2}
\sum_{{\cal E}\in B(p'^{(k)}_1, p'^{(l)}_2)} 
G_{N_1N_2}({\cal E}) \Delta {\cal E}, 
\\
\left.
\frac{d\sigma_{N_1N_2}}{d T_\mu d \Omega_\mu d\cos\theta'_1 d\cos\theta'_2}
\right|_{\cos\theta'^{(k)}_1 ,  \cos\theta'^{(l)}_2}   
&=& 
\frac{1}{\Delta \cos\theta'_1\, \Delta \cos\theta'_2}
\sum_{{\cal E}\in B(\cos\theta'^{(k)}_1,\cos\theta'^{(l)}_2)} 
G_{N_1N_2}({\cal E}) \Delta {\cal E}, 
\\
\left.
\frac{d\sigma_{N_1N_2}}{d T_\mu d \Omega_\mu d\phi'_1 d\phi'_2}
\right|_{\phi'^{(k)}_1 ,   \phi'^{(l)}_2}   
&=& 
\frac{1}{\Delta \phi'_1\, \Delta\phi'_2}
\sum_{{\cal E}\in B(\phi'^{(k)}_1,\phi'^{(l)}_2)} 
G_{N_1N_2}({\cal E}) \Delta {\cal E}, 
\\
\left.
\frac{d\sigma_{N_1N_2}}{d T_\mu d \Omega_\mu dp'_i d\cos\theta'_j}
\right|_{p'^{(k)}_i ,   \cos\theta'^{(l)}_j}   
&=& 
\frac{1}{\Delta p'_i\, \Delta\cos\theta'_j}
\sum_{{\cal E}\in B(p'^{(k)}_i,\cos\theta'^{(l)}_j)} 
G_{N_1N_2}({\cal E}) \Delta {\cal E}, 
\\
\left.
\frac{d\sigma_{N_1N_2}}{d T_\mu d \Omega_\mu dp'_i d\phi'_j}
\right|_{p'^{(k)}_i ,   \phi'^{(l)}_j}   
&=& 
\frac{1}{\Delta p'_i\, \Delta\phi'_j}
\sum_{{\cal E}\in B(p'^{(k)}_i,\phi'^{(l)}_j)} 
G_{N_1N_2}({\cal E}) \Delta {\cal E}, 
\\
\left.
\frac{d\sigma_{N_1N_2}}{d T_\mu d \Omega_\mu d\cos\theta'_i d\phi'_j}
\right|_{\cos\theta'^{(k)}_i ,   \phi'^{(l)}_j}   
&=& 
\frac{1}{\Delta \cos\theta'_i\, \Delta\phi'_j}
\sum_{{\cal E}\in B(\cos\theta'^{(k)}_i,\phi'^{(l)}_j)} 
G_{N_1N_2}({\cal E}) \Delta {\cal E}. 
\\
\end{eqnarray}

\subsection{Relative semi-inclusive cross sections}

\begin{figure}
\includegraphics[width=10cm, bb=140 430 540 760]{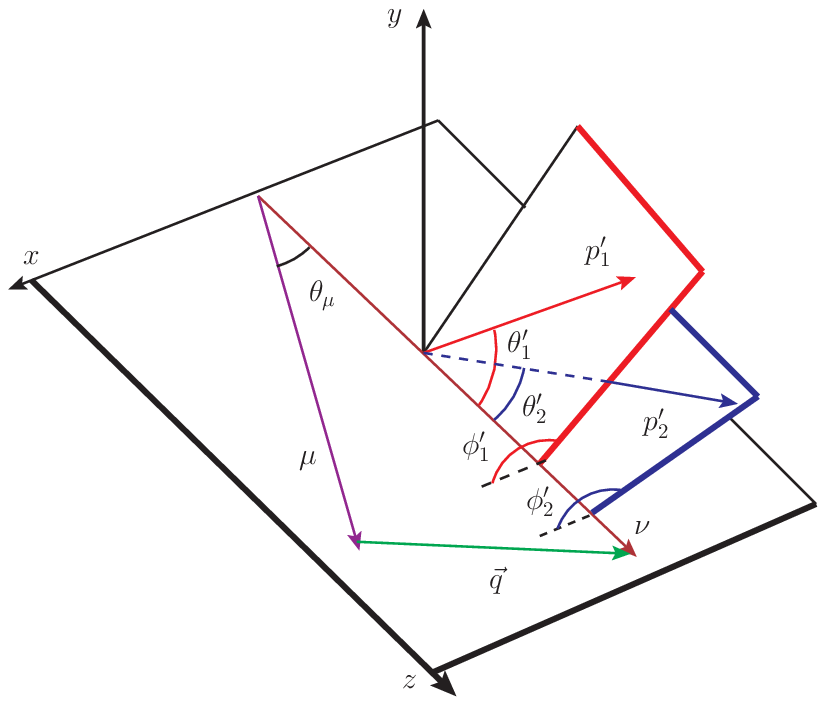}
\caption{ Kinematics for 2p2h semi-inclusive events used in this
  work. The incident neutrino defines the $z$ axis, and the muon
  direction define the scattering plane. The two final nucleon momenta
  define with the z-axis the two reaction planes with angles $\phi'_1$
  and $\phi'_2$ with respect to the scattering plane.  }
\label{fkin}
\end{figure}

Guided by the above procedure we can define the two-fold, semi-inclusive cross
section for fixed relative
angles of the final nucleons. We already mentioned that said cross section
is not measurable, but mathematically it is well defined and will
allow us to shed additional light on the effect of the hadronic tensor
on the distribution of semi-inclusive events, which is the goal
of this work.

Therefore, to proceed with the definition, we will consider the
variables $\cos\theta_r$, $\phi_r$ 
corresponding to the relative angles of each exclusive event,
which will also be distributed in bins or sub-intervals of 
$[-1,1]$, in
the case of $\cos\theta_r$, and $[0,2\pi]$ in the case of the $\phi_r$ variable.
The subset of exclusive events with relative angles in bins $(k,l)$ 
of the variables $(\cos\theta_r,\phi_r)$ is now 
\begin{equation}
B(\cos\theta_r^{(k)}, \phi_r^{(l)}) 
\equiv 
\left\{ {\cal E} \kern 5mm |  \kern 5mm
\cos\theta_r^{(k)}\leq \cos\theta_r({\cal E}) < \cos\theta_r^{(k+1)} \, 
\mbox{and}\,
\phi_r^{(l)}\leq \phi_r({\cal E}) < \phi_r^{(l+1)}
\right\}.
\end{equation}
The total probability that an event belong to  the bins $(k,l)$
of variables $(\cos\theta_r,\phi_r)$ is then  
\begin{equation}
P_{N_1N_2}(\cos\theta_r^{(k)}, \phi^{(l)})= 
\sum_{{\cal E}\in B(\cos\theta_r^{(k)},\phi_r^{(l)})}P_{N_1N_2}({\cal E}), 
\end{equation}
and the 
 two-fold 2p2h semi-inclusive cross section 
averaged over the bins $\cos\theta_r^{(k)},\phi_r^{(l)}$ is defined 
as in Eq. (\ref{two-fold}) 
\begin{equation}\label{sigmarel}
\left.\frac{d\sigma_{N_1N_2}}{d T_\mu d \Omega_\mu d\cos\theta_r d\phi_r}
\right|_{\cos\theta_r^{(k)},\phi_r^{(l)}}   
=
\frac{1}{\Delta \cos\theta_r \Delta \phi_r}
\sum_{{\cal E}\in B(\cos\theta_r^{(k)},\phi_r^{(l)})} G_{N_1N_2}({\cal E}) \Delta {\cal E}. 
\end{equation}
Once more we have the  result 
\begin{eqnarray}
\sum_{kl} 
\left.\frac{d\sigma_{N_1N_2}}{d T_\mu d \Omega_\mu d\cos\theta_r d\phi_r}
\right|_{\cos\theta_r^{(k)},\phi_r^{(l)}}   
\Delta\cos\theta_r \Delta \phi_r
& = &
\frac{d\sigma_{N_1N_2}}{d T_\mu d \Omega_\mu }.
\end{eqnarray}

\section{results}

The coordinate system and kinematics for the description of
semi-inclusive 2p2h reaction is shown in Fig. \ref{fkin}. We choose the
$z$-axis in the direction of the incident neutrino. The scattering
plane $(x,z)$ is defined by the final muon and the initial neutrino.
The transverse component of the muon momentum with respect to the
neutrino defines the $x$-direction.  The directions of the two final
momenta, $\np'_i$, and the $z$ axis define two corresponding reaction
planes that form angles $\phi'_1$ and $\phi'_2$, respectively, with
the scattering plane. The angles between $\np'_i$ and the $z$ axis are
$\theta'_i$.

Analyzing the semi-inclusive cross section, which is dependent upon
the six variables $(\mathbf{p}'_1,\mathbf{p}'_2)$, can be notably
intricate due to the necessity of selecting final events within a
six-dimensional space. Therefore, in  this initial exploration of the
two-nucleon emission process, we will focus on the distributions of
partial semi-inclusive events—--both one-fold and two-fold. This entails
the fixation of one or two final hadronic variables, with integration
being performed across the remaining ones.

As shown in the preceding section, the probability of such
partial events is computed through a partial summation of
exclusive event probabilities within a discretized framework. The
semi-inclusive variables $\mathbf{p}'_1$ and $\mathbf{p}'_2$ will
similarly be discretely binned. 

For the sake of convenience, we will further present the outcomes in
terms of probabilities concerning the distributions of feasible events
of both one-fold and two-fold nature, while maintaining a constant
neutrino kinematic configuration. The corresponding one-fold and two-fold
cross-sections are directly proportional to these probabilities, with
the proportionality factor being precisely the inclusive 2p2h cross-section.
 This value, established for the lepton kinematics, is
divided by the bin volume, as indicated in Equations
(\ref{probability}) and (\ref{probability3}).
In this study, we will not delve into averaging over the
neutrino flux. Instead, we will operate under the assumption of an
incident neutrino possessing a predetermined energy, yielding a muon
with certain energy and scattering angle.

\begin{figure}
\includegraphics[width=7cm, bb=120 50 440 760]{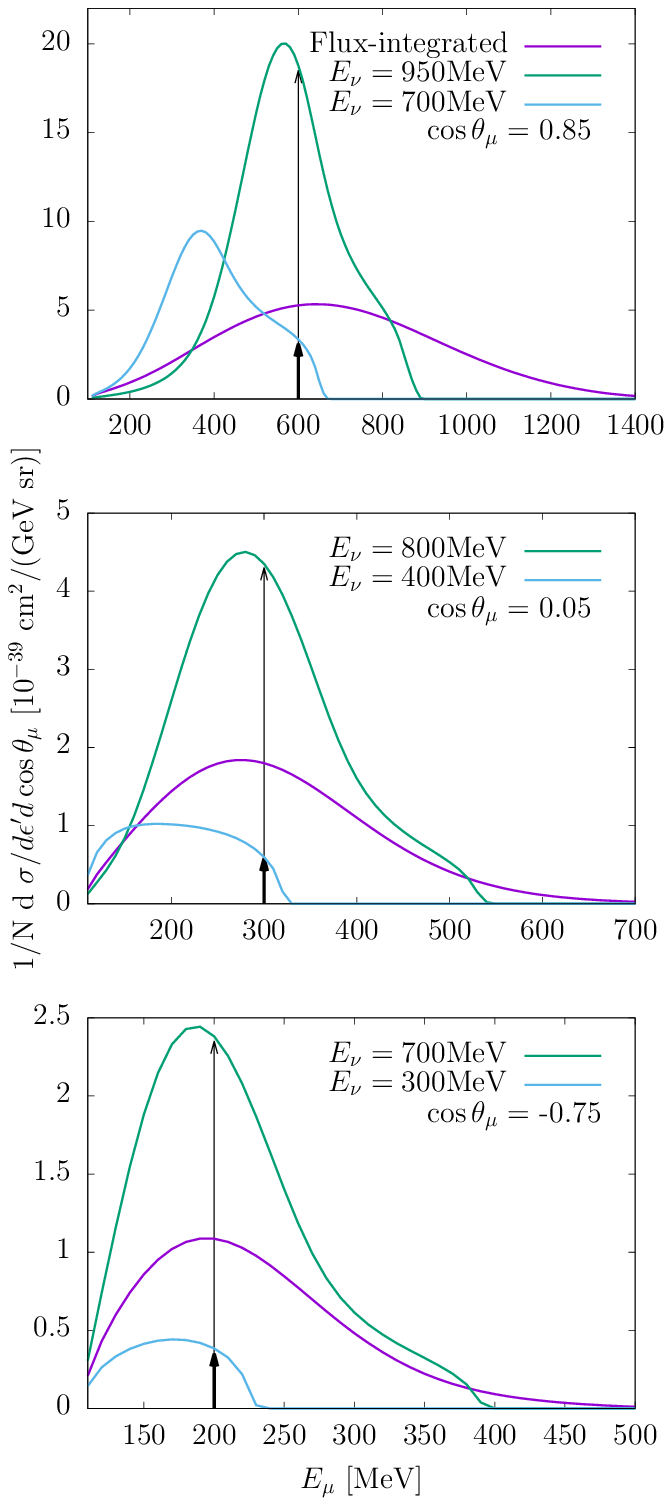}
\caption{Inclusive 2p2h neutrino cross section as a function of muon
  energy. Each panel corresponds to a fixed muon angle. In
  each panel we plot the cross section for two neutrino energies
  compared to the cross section averaged over the neutrino flux. The
  neutrino energies are those that appear in table I. The arrows
  indicate the muon energies in the table. }
\label{sfig1}
\end{figure}

\begin{table}
\begin{tabular}{cccccccc}
\hline
Kinematics & $E_\nu$ (MeV) & $E_\mu$ (MeV) & $\cos\theta_\mu$ & $\theta_\mu$ [deg] & $q$ [MeV/c] & $\omega$ [MeV] & $\theta_q$ [deg] \\ \hline \hline
1  & 950  & 600 & 0,85  & 31,79 & 545,42 & 350  & 34,78 \\ \hline
2  & 700  & 600 & 0,85  & 31,79 & 368,77 & 100  & 57,53 \\ \hline
3  & 800  & 300 & 0,05  & 87,13 & 834,49 & 500  &  19,64 \\ \hline
4  & 400  & 300 & 0,05  & 87,13 & 477,07 & 100  & 36,00 \\ \hline
5  & 700  & 200 & $-0,75$ & 138,59& 834,93 & 500  & 7,73 \\ \hline
6  & 300  & 200 & $-0,75$ & 138,59& 441,85 & 100  & 14,72 \\ \hline
\end{tabular}
\centering
\caption{\label{tabla}
Selected lepton kinematics studied in this work: neutrino
  energy, muon energy, scattering angle and the corresponding values
  of $q,\omega$ and also the angle $\theta_q$ between $\nq$ and the
  neutrino direction ($z$-axis).}
\end{table}

We will consider the six kinematics given in table I.  These
kinematics have been chosen as follows. First of all we choose three
scattering angles $\cos\theta_\mu =0.85$, 0.05 and $-0.75$
corresponding to three angular bins from the MiniBoonE experiment. In
Fig. 3 we plot the inclusive 2p2h cross section averaged with the
neutrino flux against the muon energy, for the three given angles.
The position of the maximum on this cross section is indicated with
the arrows in each panel, that corresponds approximately 
to the energy of the muon
in the kinematics of table I.  In each panel of Fig. 3 we also show
the 2p2h cross section for two fixed neutrino energies. The value of
the first neutrino energy in the green curves is chosen so that the
position of the maximum of the cross section roughly coincides with
that of the averaged cross section. This is indicated by long thin
arrows.  These neutrino energies corresponds to kinematics 1, 3 and 5
of tab. I,

 The value of the second neutrino energy in the blue curves of fig. 3 is
chosen such that $\omega=100$ MeV at the maximum of the average cross
section. This corresponds to the kinematics 2, 4 y 6 of Tab. I.
A short thick arrow shows the contribution of this second
neutrino energy to the cross section before averaging in the flux.
These neutrino energies  give a  smaller
contribution to the 2p2h cross section.

Once the lepton kinematics are defined, the lepton tensor can be immediately
calculated with Eq. (\ref{leptonic}) taking into account that the
leptonic vectors and momentum transfer in the coordinate system of
Fig. 2 are the following
\begin{eqnarray}
k^\mu=(E_\nu,\nk ) &=& (E_\nu,0,0,E_\nu)\\
k^\nu=(E_\mu,\nk') &=&(E_\mu,k'\sin\theta_\mu,0,k'\cos\theta_\mu)\\
\nq=\nk -\nk'  &=& (-k' \sin\theta_\mu,0,E_\nu-k' \cos\theta_\mu)
\end{eqnarray}

As explained in the last section, for each kinematic we generate
exclusive events in terms of six coordinates
$(h_1^3,\cos\theta_1,\phi_1,h_2^3,\cos\theta_2,\phi_2)$ for the two
holes, and two relative angles $(\cos\theta_r,\phi_r)$ for the final particles.
We generate a
number of $7^6$ hole pairs and $200^2$ pairs of relative angles.  The
total number of exclusive events, ${\cal E}=(\nh_1,\nh_2,\np'_1,\np'_2)$,
generated in this way is $N_{\rm events}=4.71\times 10^9$ for each one
of the kinematics of Tab. I.

Now, for each exclusive event, ${\cal E}_i$, we compute the exclusive
hadronic tensor $w^{\mu\nu}_{N_1N_2}({\cal E}_i)$ for PP and NP (NN
and NP) ejection in the case of neutrino (antineutrino)
scattering. After contraction with the leptonic tensor, we build the
values of the function $G_{N_1N_2}({\cal E}_i)$ of Eq. (\ref{gnn}) that
determine the probability distribution of events and the
semi-inclusive cross sections. We sum over the events included in the selected
bins, and compute the corresponding one-fold and two-fold
semi-inclusive cross sections averaged over the bin using the equations
(\ref{one-fold}) and (\ref{two-fold}).  Dividing by the inclusive
cross section, the probability distribution of partial semi-inclusive
events is obtained.

In this work we are interested in comparing with the pure phase-space
 isotropic distribution of final-state nucleons in the hadronic
center-of-mass frame, similar to what is done in the Monte Carlo
implementations \cite{Sob12}. 
This is equivalent to neglecting the dependence of the hadronic tensor
$W^{\mu\nu}_{N_1N_2}({\cal E})$
 on the exclusive event ${\cal E}$.  
In our calculation we simply set
\begin{equation}
\sigma_0 L_{\mu\nu}W^{\mu\nu}_{N_1N_2}({\cal E})=1.
\end{equation}
The resulting semi-inclusive event distribution is only due to the
kinematics of 2p2h phase-space (PS) and does not depend on the current
matrix elements. In this case the probability of exclusive events is computed 
similarly to Eq. (\ref{probE})
\begin{equation} \label{probF}
P_{\rm P.S.}({\cal E}) = \frac
 { G_{\rm P.S.}({\cal E}) }
 { \sum_{\cal E'} G_{\rm P.S.}({\cal E'})}
\end{equation}
where the exclusive 
phase-space $G_{\rm P.S.}$ function is defined similarly to Eq. (\ref{gnn})  
\begin{equation}  \label{g0}
G_{\rm P.S.}(\nh_1,\nh_2,\theta_r,\phi_r)
=\frac{V}{(2\pi)^9}
\frac{(m^*_N)^4}{E_1E_2}
\frac{p_1''}{2E_1''}
\theta(p'_1-k_F)\theta(p'_2-k_F)
\theta(E'^2-p'^2-4m_N^{*2})
\end{equation}
Note that the $G_{\rm P.S.}$ function do not depend on the charge of the
nucleons,  it only depends on the phase-space kinematics.  
Using this phase-space distribution  we can define 
a semi-inclusive cross section 
whose integral is equal to the inclusive cross-section,
similarly to Eq. (\ref{probability2})
\begin{equation} \label{probability2}
\left.
\frac{d\sigma_{\rm P.S.}}{ d T_\mu d \Omega_\mu dX dY}
\right|_{X^{(k)},Y^{(l)} }   \equiv
 \frac{P_{\rm P.S.}(X^{(k)},Y^{(l)})}{\Delta X \Delta Y} 
\left(\frac{d\sigma}{d T_\mu d \Omega_\mu}\right)_{2p2h}
\end{equation}

\begin{figure}
\includegraphics[width=8cm, bb=190 40 440 720]{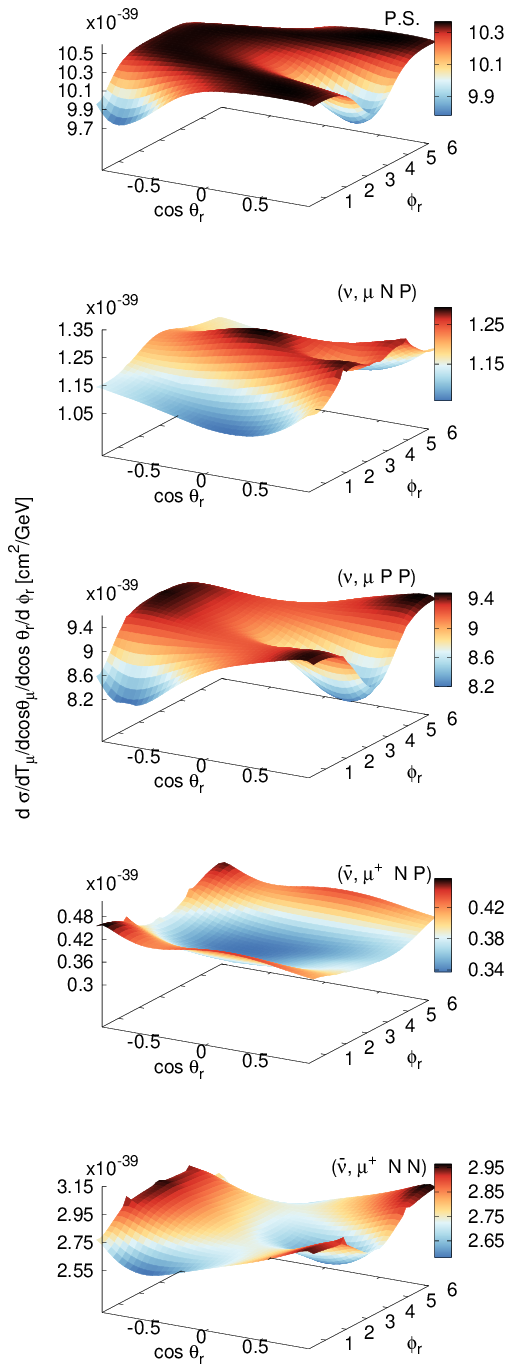}
\caption{Two-fold Semi-inclusive cross section 
with respect to the relative angles 
in the center of mass system of the two outgoing nucleons. 
From top to bottom we show the phase-space, NP and PP  emission with neutrinos,
and NP and NN emission with antineutrinos. The kinematics is \#1 of Tab. 
\ref{tabla}.  }
\label{fig4}
\end{figure}

\begin{figure}
\includegraphics[width=8cm, bb=190 40 440 720]{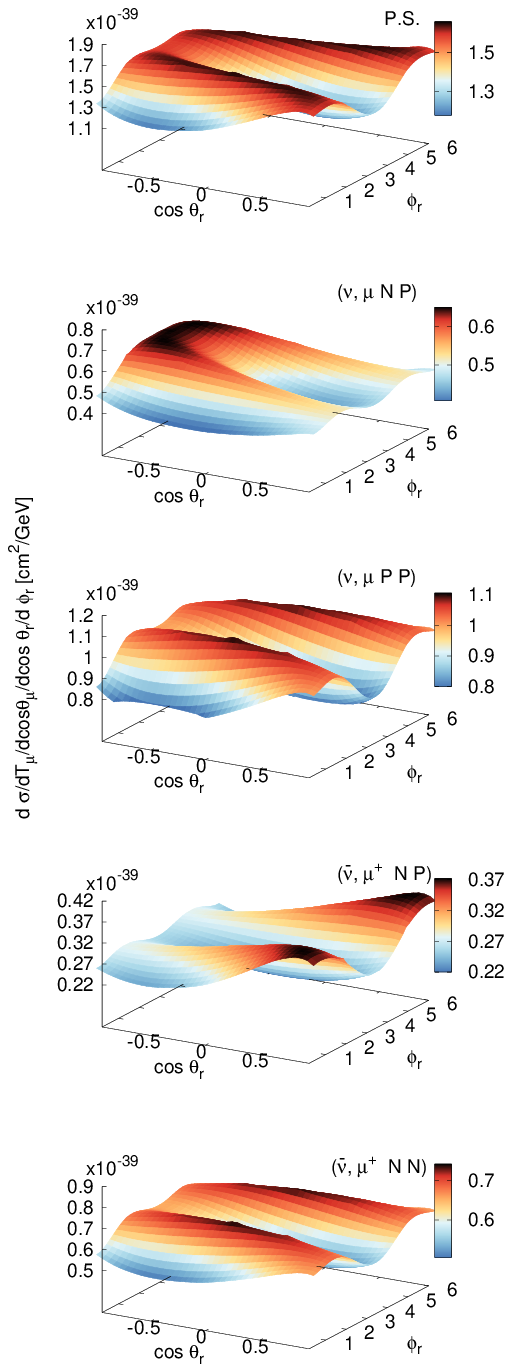}
\caption{The same as Fig. \ref{fig4} for the kinematics \#2 of
  Tab. \ref{tabla}.    }
\label{fig5}
\end{figure}

\subsection{Two-fold distributions}

In Figs. \ref{fig4} and \ref{fig5} we show our results for the
two-fold semi-inclusive cross section of the relative angles in the
center of mass system of the two outgoing nucleons, given by
Eq. (\ref{sigmarel}), for kinematics 1 and 2 of Table \ref{tabla}.
In the top panels we show the phase-space results. 
The NP and PP channels for neutrino scattering, $(\nu_\mu,\mu NP)$,
and $(\nu_\mu,\mu PP)$, respectively, 
are also shown separately, as well as the NP and
NN channels for antineutrino scattering, $(\overline{\nu}_\mu,\mu^+
NP)$, and $(\overline{\nu}_\mu,\mu^+ NN)$, respectively.  All of them have been
calculated including the full 2p2h hadronic tensor (\ref{elementary}).

To begin our analysis, we first verified that the integrals of the
distributions, as shown in Figs. 4 and 5 over the relative angles,
yield the value of the inclusive 2p2h cross section for the specified
kinematics and charge channel. This serves as a valuable test
of our present calculation. In refs.
\cite{Mar21,Mar21b,Mar23,Mar23b}, inclusive 2p2h response functions
were computed within a reference frame with the momentum transfer
pointing along the $z$-axis, utilizing a seven-dimensional integration
scheme. In contrast, in the present calculation the $z$-axis is
aligned with the neutrino direction, necessitating an
eight-dimensional integration. Moreover, we compute all components of
the hadronic tensor, while in Refs. \cite{Mar21,Mar21b,Mar23,Mar23b}
only the inclusive response functions where calculated.  This
cross-check reinforces the consistency of our results.

In  Figs. 4 and 5 the phase-space (PS) distributions are appropriately
normalized to the total neutron-proton (NP) plus proton-proton (PP)
inclusive neutrino cross section. A close inspection reveals that
indeed, in the case of neutrino scattering, the sum of the NP and PP
cross sections approximately matches the PS cross section in the upper
panel. Furthermore, it is evident that the PP cross section is
approximately seven times greater than the NP cross section.  A
similar value for the PP/NP ratio was reported in \cite{Rui16b} around
the $\Delta$ peak of the 2p2h inclusive response.

It is noteworthy that the NP and PP distributions differ from the
phase-space distribution. This discrepancy arises from our inclusion
of the exact dependence of the hadronic tensor on the exclusive
variables of 2p2h excitations. It is noteworthy that the PS
distribution is not perfectly uniform, primarily due to the summation
over holes. Consequently, the center-of-mass momentum
is not constant, resulting in an averaged distribution over all
possible values of the center of mass contributing to the
semi-inclusive observable.  According to Table I, the results of Fig. 4
correspond to neutrino energy $E_\nu=950$ MeV, $q=545$ MeV/c and
$\omega=350$ MeV. For this case it is apparent that according to phase
space, the most probable events cluster around the darker continuous
band show in the top panel.  However, a noticeable difference is
observed compared to the neutrino PP and NP distributions, with the
differences being more pronounced in the case of NP
emission. Additionally, we note that the distribution peaks for NP and
PP emission occur at different angles. In the case of antineutrinos, distinct
distributions are also observed for NP and NN emissions, which deviate
from the phase-space predictions. These observations emphasize the
importance of considering the full dependence of the hadronic tensor
on the exclusive variables.

\begin{figure}[b]
\includegraphics[width=14cm, bb=60 185 520 730]{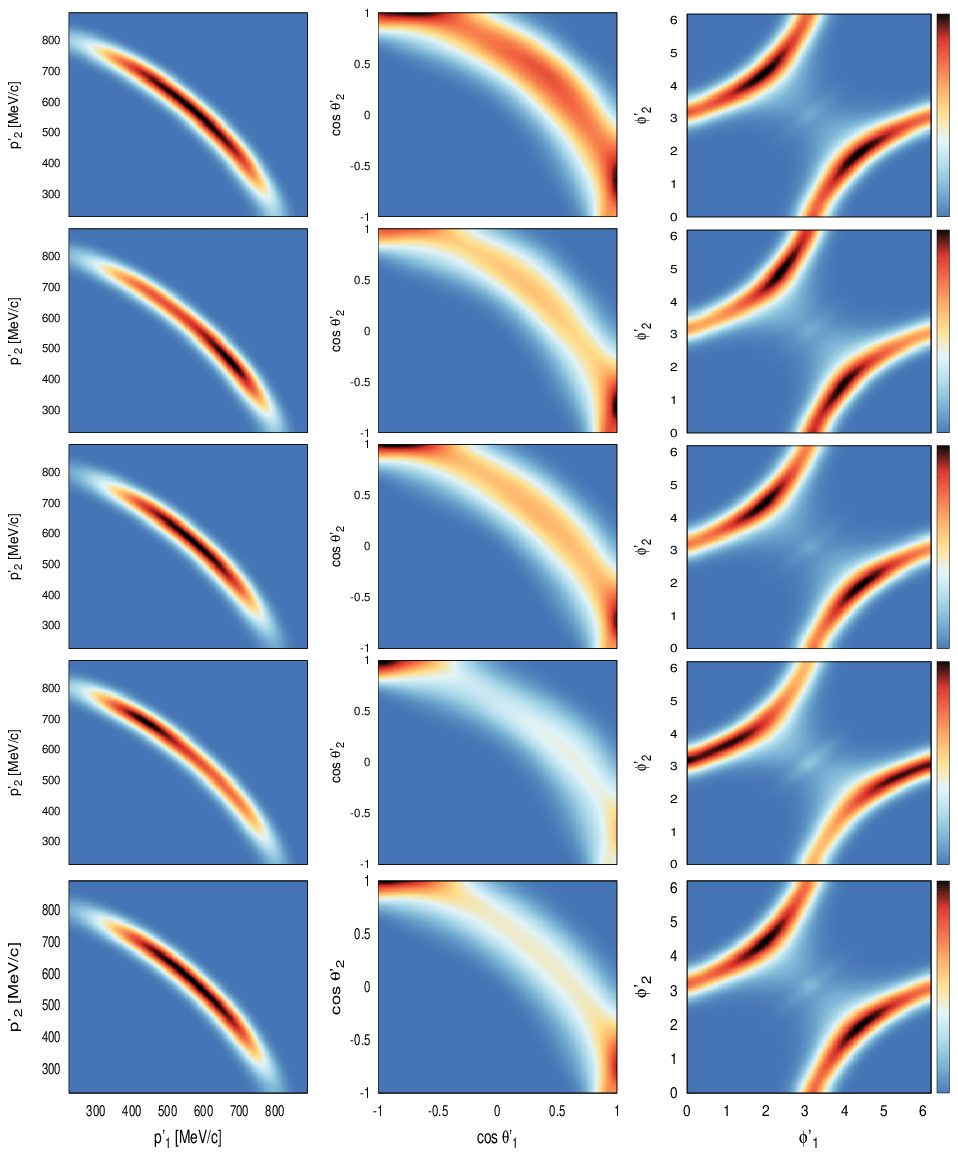}
\caption{ Two-fold distributions, namely $P(p_1,p_2)$,
  $P(\cos\theta_1,\cos\theta_2)$, and $P(\phi_1,\phi_2)$, for
  kinematics \#1.  From top to bottom we show the phase space, the
  neutrino NP and PP, and the antineutrino NP and NN emission
  channels, respectively. The units in each panel are such that 
the corresponding distribution is normalized to one.
}
\label{sfig3}
\end{figure}

The distributions of the relative angles also exhibit a significant
dependence on the leptonic kinematics, as can be discerned when
comparing them with the case of different kinematics, as exemplified
in Fig. 5. In this scenario, characterized by a neutrino energy of 700
MeV, momentum transfer $q=369$ MeV/c and energy transfer $\omega= 100$
MeV, notable distinctions become apparent.  In the case of neutrino
scattering, the PP distribution closely aligns with the pure phase
space, a feature corroborated by the location of the darker region
where the cross section is larger. Conversely, the NP distribution
deviates considerably from both the phase space and the PP
distribution, with its maximum found at backward angles
$(\theta_r)$. In this instance, the NP cross
section is slightly smaller than that of the PP, although it remains
of the same order of magnitude.

Similar trends are observed in the case of antineutrino scattering in Fig. 5.
Turning our attention to the antineutrino NN cross section, it closely
resembles the phase space distribution, whereas the NP cross section
exhibits noticeable differences. In this case, the maximum is shifted
towards forward angles in $\theta_r$. These nuanced variations
underscore the intricacies of the semi-inclusive reactions and
emphasize the impact of the hadronic tensor and lepton kinematics on
the semi-inclusive 2p2h cross section.

The results presented in Figures 4 and 5 are expressed in terms of
absolute cross-section values. We have observed similar trends,
demonstrating deviations of the semi-inclusive distributions from
those expected for pure phase space, for all the leptonic kinematics
of Tab. I. Notably, these differences are most pronounced for PN
emission.
These results provide a partial integration
of the total semi-inclusive cross section. To gain a broader view of
the entire six-dimensional landscape but in two-dimensional, averaged
sections, we will now display results for the two-fold cross sections
involving pairs of the observable semi-inclusive 
variables $(p'_i,\cos\theta_i,\phi_i)$.
In this context, it is less critical to ascertain the absolute value
of the cross section, as we are aware that the normalization is
uniform across all distributions, with the inclusive cross section
serving as a common reference for a given kinematic
scenario. Therefore, we will present the results in terms of
probabilities for each semi-inclusive bin, understanding that the
summation over bins in these distributions equals one. Our primary
objective is to examine the differences among various semi-inclusive
charge channels and their distinctions relative to a pure phase-space
distribution.

We show a more complete example of the available two-fold
distributions in Figs. 6, 7 and 8.  In the interest of brevity and due
to space limitations, we have chosen to focus on representative
examples of these combinations, mindful of not overwhelming the reader
with an exhaustive display of every possible distribution. It is
essential to acknowledge that illustrating every conceivable
combination is impractical within the confines of this
presentation. However, the selected distributions provide a meaningful
and insightful glimpse into the complex multi-dimensional landscape of
semi-inclusive reactions initiated by neutrinos and antineutrinos.
Together, these figures provide a comprehensive perspective on the
distributions within the semi-inclusive charge
channels, enabling a deeper understanding of the deviations from pure
phase-space models.

As a first example, in Figure 6, we present the two-fold
distributions, namely $P(p'_1,p'_2)$, $P(\cos\theta'_1,\cos\theta'_2)$,
and $P(\phi'_1,\phi'_2)$, for both neutrino and antineutrino
scattering. These distributions provide insights into the joint
probability distributions of pairs of momenta or emission angles of
the two particles within the semi-inclusive reactions.  This example
vividly demonstrates the correlations between pairs of variables and
the differences introduced by considering the semi-inclusive hadronic
tensor in contrast to pure phase-space distributions. Notably, these
distributions show the asymmetries present
in the final-state momenta of protons and neutrons within the NP
channel, underscoring the influence of the hadronic tensor.

Specifically, the $P(p'_1,p'_2)$ distribution (first column of Fig. 6)
reveals a strong correlation between these two variables. The
distribution  occupies a narrow band around an
curve centered at \((k_F,k_F)\), corresponding to the
minimum values the outgoing particles can attain. The global
shape of the distribution is primarily dictated by kinematics, 
an does not depend on the hadronic tensor.
We observe that the distribution is centered and exhibits a
maximum around emission momenta of $p'_1 = p'_2 \sim 550$ MeV/c for PP, NN, and
the phase-space (PS) distributions. However, the NP distribution
reveals a distinct pattern. For neutrino scattering, its maximum is
centered approximately at $p'_1 \sim 650$ MeV/c and 
$p'_2 = 450$ MeV/c, while for
antineutrino scattering, the situation is reversed, bearing in mind
that particle 1 is a neutron and particle 2 is a proton. This clear
asymmetry underscores the dissimilarity in final proton and neutron
momenta within the NP channel, a consequence of the influence of the
hadronic tensor.

In the second column of Figure 6, we turn our attention to the
correlations between the emission angles $\theta'_1$ and $\theta'_2)$.
Here we also observe a distinctive pattern: the emission angles tend to
cluster within a relatively narrow band, forming a curve that
traverses from backward-forward $(-1, 1)$ to forward-backward $(1,
-1)$ emissions. Within this band, all pairs of angles exhibit fairly
similar probabilities, except in the extremes.  Additionally,
asymmetries in the distributions of NP emission concerning particles 1
and 2 are observed, even though they might not be as evident in the
graph. Meanwhile, the distributions for PP, NN, and PS 
exhibit clear symmetry under the $1\leftrightarrow 2$ exchange.

In the third column of Figure 6, we finally explore the correlations between
the azimuthal emission angles $\phi'_1$ and $\phi'_2$, which correspond to the
angles between the emission plane and the scattering plane (as
illustrated in Fig. 2). Once again, we observe a distinct correlation
shape that traces two precise trajectories in the 
$(\phi_1,\phi_2)$ plane,
ranging from $(0,\pi)$ to $(\pi,2\pi)$ and from 
$(\pi,0)$ to $(2\pi,\pi)$. It's worth
noting that in spherical coordinates, the values 0 and $2\pi$ are
identified as the same point, so these two trajectories actually form
a single path.
In this case, we observe a clear $1\leftrightarrow2$ 
symmetry in the distributions for PP,
NN, and the phase space around the angles of $(\pi/2,3\pi/2)$ and
$(3\pi/2,\pi/2)$. However, in the case of NP emission, this symmetry is
disrupted again due to the influence of the hadronic tensor.

\begin{figure*}
\includegraphics[width=12cm,  bb=60 280 520 730]{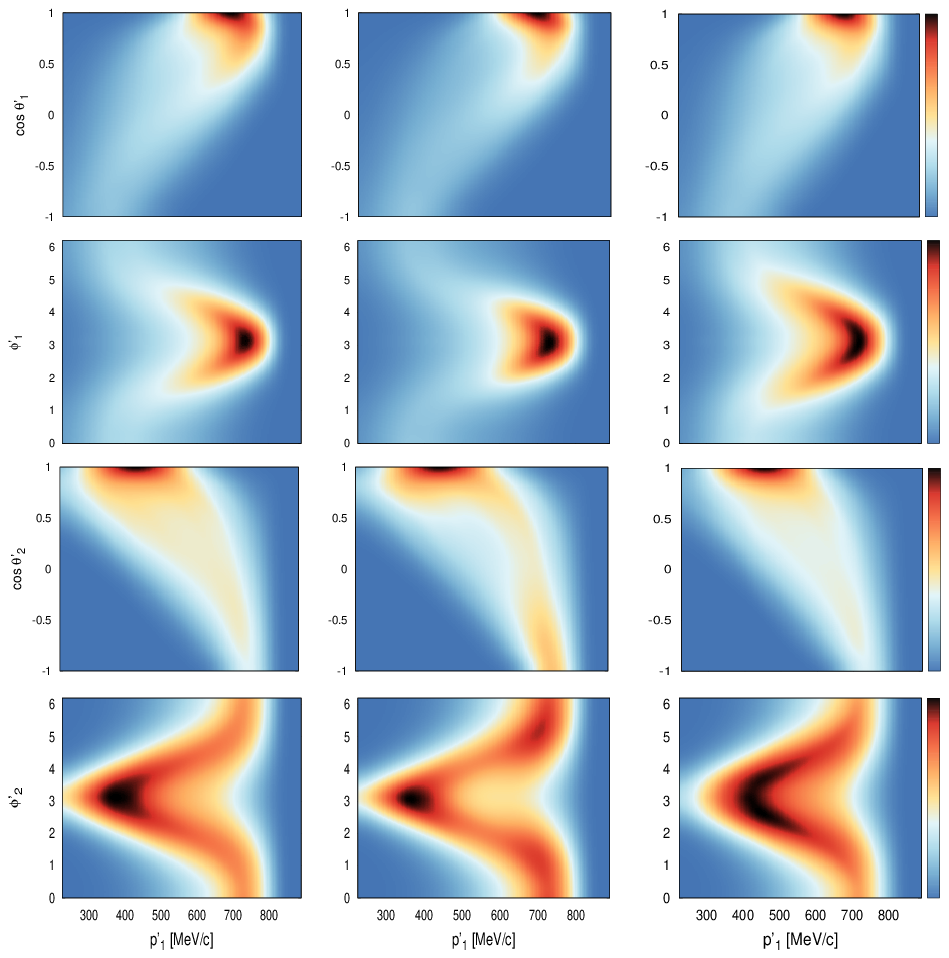}
	\caption{ Two-fold distributions, $P(p_1,\cos\theta_1)$,
          $P(p_1,\phi_1)$, $P(p_1,\cos\theta_2)$, and $P(p_1,\phi_2)$,
          specifically for neutrino scattering and kinematics 1.  In
          each panel from left to right we show the phase space, NP
          and PP distributions.  }
\label{sfig4}
\end{figure*}

As a second example, depicted in Figure 7, we present the two-fold
distributions $P(p'_1,\cos\theta'_1)$, $P(p'_1,\phi'_1)$,
$P(p'_1,\cos\theta'_2)$, and $P(p'_1,\phi'_2)$, specifically for
neutrino scattering and kinematics 1.  These distributions shed light
on the correlations between momentum of the first particle and one of
the four angular variables.  When comparing the phase-space, NP, and
PP distributions, it is observed that they share a similar shape. This
similarity is attributed to the fact that, as in the previous cases,
the overall features are largely determined by kinematics,
specifically energy and momentum conservation. The influence of the
hadronic tensor alters the finer details, and clear differences can be
observed between NP and PP emissions when compared to the phase-space
distribution.

\begin{figure*}
\includegraphics[width=16cm, bb=60 500 520 730]{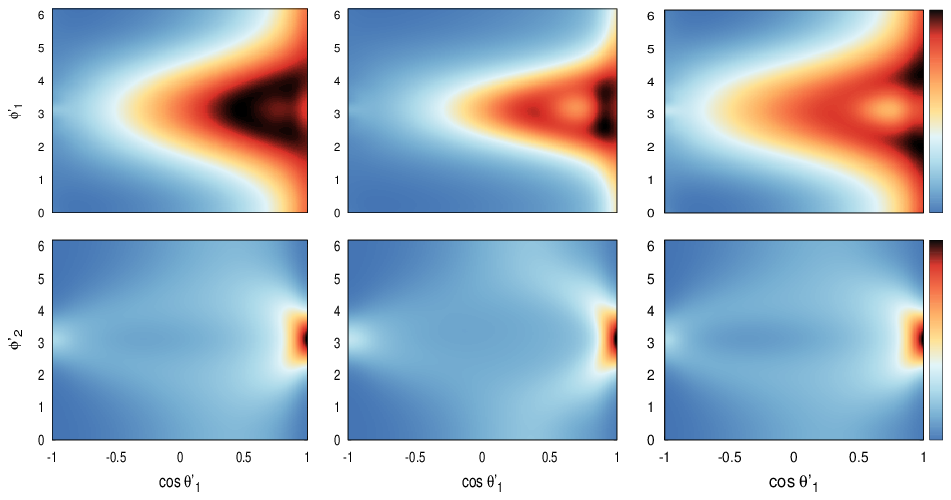}
\caption{Two-fold semi-inclusive distributions, 
$P(\cos\theta'_1,\phi'_1)$
and $P(\cos\theta'_1,\phi'_2)$, 
for neutrino scattering and for kinematics 1. From left to right we show the phase-space, NP and PP emission channels, respectively.
}
	\label{sfig5}
\end{figure*}

In a final example, in Figure 8, we showcase the two-fold
distributions, $P(\cos\theta'_1,\phi'_1)$ and
$P(\cos\theta'_1,\phi'_2)$, for neutrino scattering and kinematics 1,
highlighting the joint probabilities associated with the emission
angle of particle one and one of the azimuthal angles in the
semi-inclusive reaction.  Once more, we observe that the overall
shapes of these distributions are similar among the three cases:
phase-space, NP, and PP emissions. However, notable differences become
evident in the detailed behavior of $P(\cos\theta'_1,\phi'_1)$,
indicating that this distribution is particularly sensitive to the
influence of the hadronic tensor.  The distribution of
$\cos\theta'_1$ and $\phi'_2$ exhibits a pronounced peak around
forward emission, and it becomes smoother away from that region. While
there are still differences between the various emission channels,
these differences are not very noticeable in the figure.

\begin{figure*}
\includegraphics[width=12cm, bb=60 160 510 730]{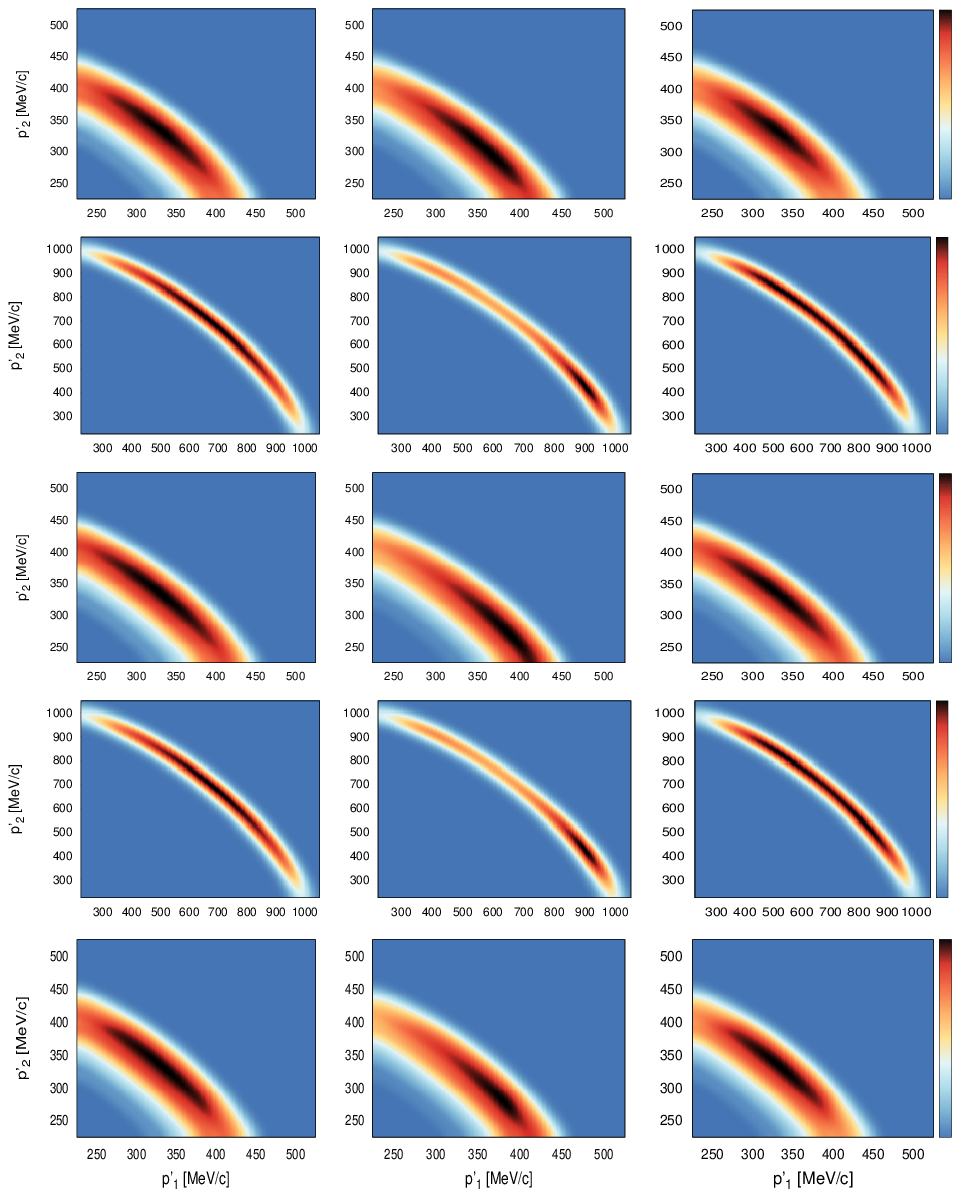}
	\caption{ Two-fold distribution $P(p'_1,p'_2)$ for
          semi-inclusive 2p2h neutrino scattering. From top to bottom,
          the rows refer to the kinematics 2--6 of table I, Form left
          to right, the columns refer to the phase-space, NP and PP
          distribution, respectively.  }
	\label{sfig6}
\end{figure*}

To conclude the discussion of two-fold distributions in semi-inclusive
two-nucleon emission, Figure 9 provides a comprehensive comparison of
the distribution $P(p'_1,p'_2)$.  This comparison covers kinematics 2
to 6 (rows in Fig. 9) for neutrino scattering, as outlined in Table
I. The three columns, from left to right, correspond to phase-space,
NP, and PP distributions, respectively.  In this analysis, a clear
correlation between the two momenta remains evident across all
cases. In the instances of phase-space and PP emissions, the
distributions exhibit clear symmetry with respect to
$1\leftrightarrow$ exchange.  Then $p'_1$ equals $p'_2$.  at the
maximum of the distribution.  Conversely, the NP distribution displays
a clear asymmetry under $1\leftrightarrow$ exchange, and $p'_1\ne
p'_2$ at the maximum. These result suggest that it is more likely for
the neutron to possess more energy than the proton.
The consistent trend of distinct distributions of momentum pairs
between the NP and PP channels is observed across all the kinematic
configurations analyzed in this study. This consistency underscores
the necessity of including the hadronic tensor to accurately account
for these differences.

\begin{figure*}
	\includegraphics[width=11cm, bb=110 80 520 730]{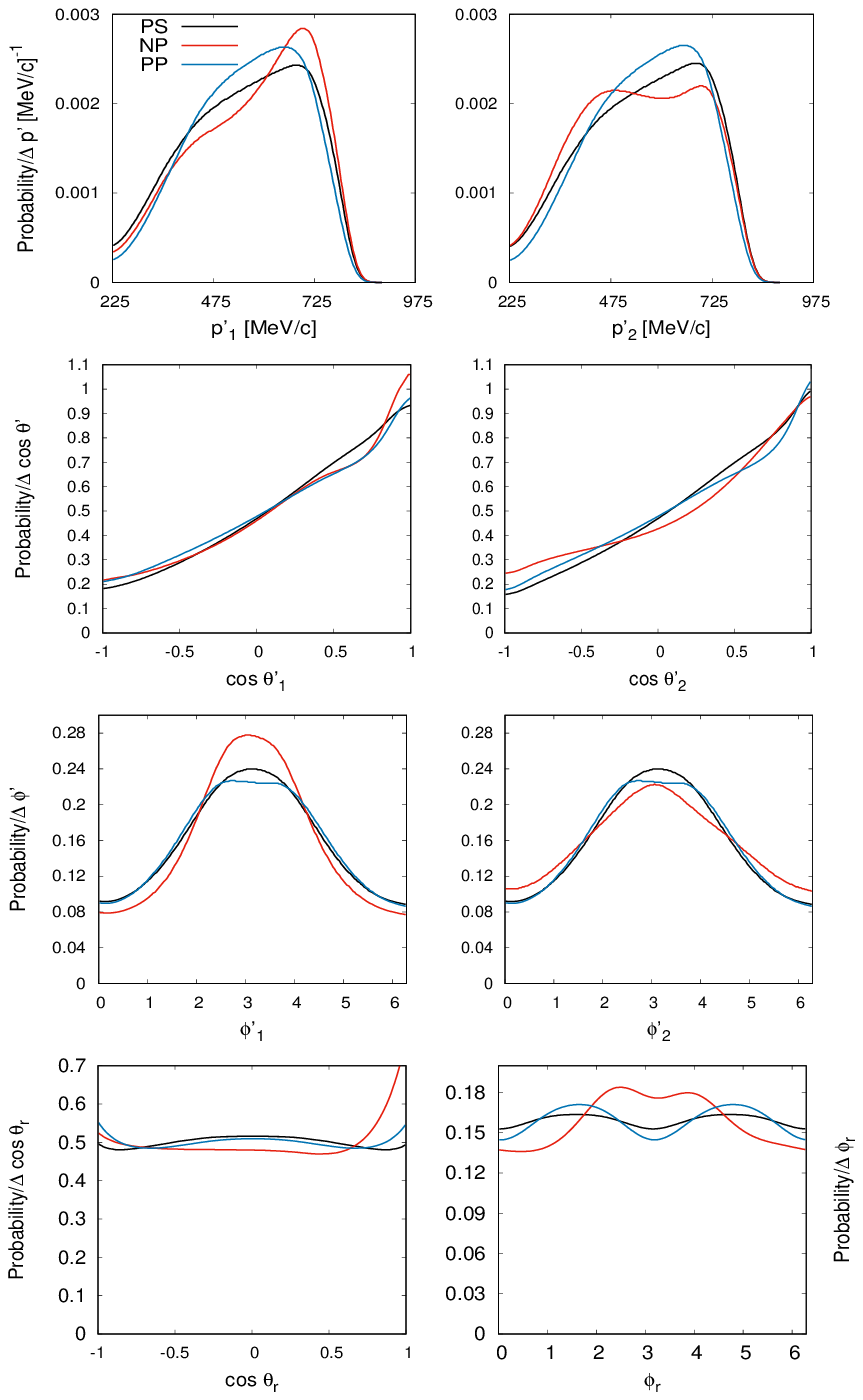}
	\caption{ 
One-fold distributions for semi-inclusive
          two-nucleon emission induced by neutrinos for kinematics
          1. In each panel, we compare the distributions for
          phase-space (PS), NP, and PP emissions.
          }
	\label{sfig7a}
\end{figure*}

\subsection{One-fold distributions}

Now we discuss the results for the one-fold distributions in
semi-inclusive two-nucleon emission induced by neutrinos
(antineutrinos).  Note that these one-fold cross sections also
contribute to semi-inclusive one-nucleon emission when combined with
the results obtained from a model of one-nucleon knockout. 

  In Figure 10, we present the distributions specifically for neutrino
  scattering, focusing on kinematics 1 as outlined in Table I. In each
  panel, we display the one-fold distributions of the variables $p'_1,
  p'_2, \cos\theta'_1, \cos\theta'_2,\phi'_1,\phi'_2$ as well as the
  angles in the center-of-mass (CM) frame of the final particles
  $\cos\theta_r$ and $\phi_r$. Within each panel, we provide a
  comparison between the distributions for phase-space (PS), NP, and
  PP channels.

In these distributions we fix the bins for one variable and integrate
with respect to all other variables (or sum over all remaining bins).
As a consequence  the one-fold distributions  on the different
channels appears quite similar. This is, in a way, an averaging effect
on the hadronic tensor.  However, despite this overall similarity, the
NP channel exhibits notable distinctions from the NP and PP channels.
The differences between the phase space and the two emission channels
are typically on the order of 10\%--20\% depending on the kinematics.  

We observe, when comparing the two top panels, that the distribution
$P(p'_1)$ is equal to the $P(p'_2)$ distribution for PP emission. This
equality arises due to the symmetry under the exchange of two
protons. The same symmetry holds for the phase space, but in this
case, it's because, by definition, the phase-space distribution
depends solely on kinematics. However, in the case of NP emission,
this symmetry is not observed due to the inherent differences between
the two outgoing particles and the isospin dependence of the hadronic
tensor. Recall that particle 1 is a neutron and particle 2 is a
proton. Consequently, the neutron distribution exhibits a more
pronounced peak at higher momenta compared to the proton distribution.

Moving on to the panels of the second row in Fig. 10, we examine the
distributions with respect to the polar emission angles,
$P(\cos\theta'_1)$ and $P(\cos\theta'_2)$. Ideally, these two distributions
should be exactly equal for the phase space, but in the figure, they
do not appear exactly identical one being slightly higher than the
other. This small discrepancy results from the numerical error
introduced when discretizing the integrals into bins and the
calculation method not treating particles 1 and 2
symmetrically. Consequently, the differences between these curves
provide an estimate of the numerical error incurred when calculating
these angular distributions, which here is approximately 5\%. 
To reduce this error,
one would need to decrease the bin size accordingly. However, this
would significantly increase the number of exclusive events that need
to be summed (and computed). It's important to keep this error in mind
when interpreting the results of this study.
Looking at the probability distributions in the panels of the second
row, we notice that they increase with $\cos\theta'_i$, which means they
decrease with the angle itself. This indicates that forward emission
(with forward being the direction of the neutrino) is more probable,
while backward emission is less likely.

Now we examine the two panels of the third row in Fig. 10 that display
the distributions over the azimuthal angles. This indicates the
probability of the orientation of the reaction plane of particle one
with respect to the scattering plane when the other emerges in any
direction, or vice versa in the case of particle 2. We observe that
the prevailing trend is to be emitted predominantly in the semi-plane
with an angle of $\pi$, which is where the momentum transfer vector is
contained (see Fig. 2). There is an asymmetry in the case of NP emission when
switching particle 1 for particle 2. This results in the neutron
(particle 1) having a greater tendency to be emitted in the $\pi$-plane (and
adjacent planes) than the proton (particle 2).

Let's now examine the distributions with respect to the emission
angles in the CM of the two final particles, shown in
the bottom panels of Fig. 10. The distribution of $\cos\theta_r$ is
quite flat in the case of phase space and PP emission, indicating that
PP emission is approximately consistent with an isotropic distribution
in the center of mass. However, it's also evident that NP emission is
not as compatible with this hypothesis, as the angular distribution
deviates from the phase space distribution. It is larger for forward
angles and smaller for other angles.

In the case of the distribution of the relative azimuthal angle
$\phi_r$, both the PP and NP channels deviate form a pure isotropic
distribution in the CM.  The NP channel stands out as distinctly
different from the other two. In this case, both the distributions for
the PS and PP channels exhibit two maxima around $\pi/2$ and $3\pi/2$,
with a minimum at $\pi$.  In contrast, the NP distribution displays
two larger maxima that are closer to $\pi$.

\begin{figure*}
	\includegraphics[width=11cm, bb=110 80 520 730]{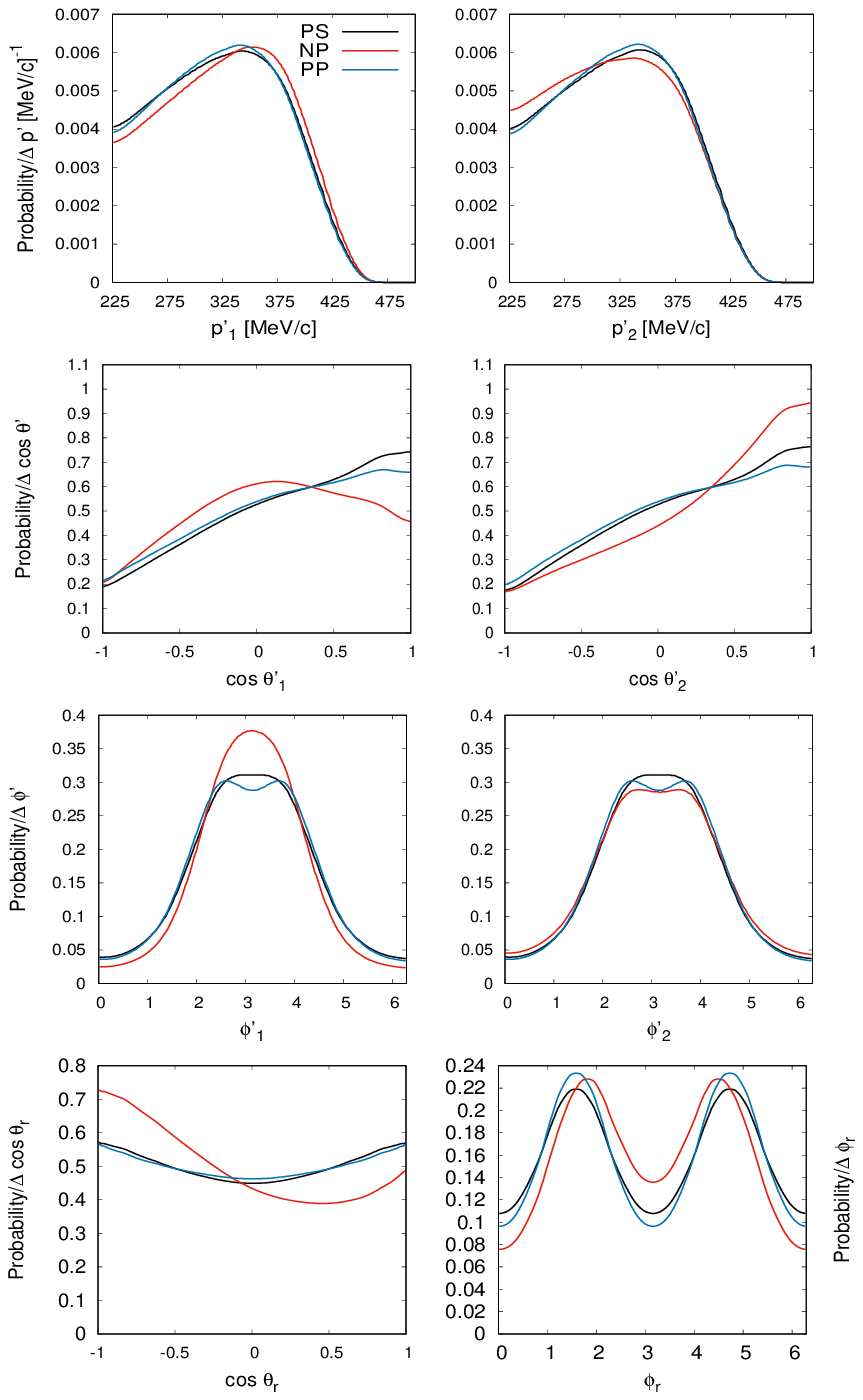}
	\caption{The same of Fig. 10 for kinematics 2.}
	\label{sfig7b}
\end{figure*}

Similar differences are observed in Figure 11, which corresponds to
kinematics 2. However, comparing with Figure 10, it becomes evident
that the specific shape of the distribution also whimsically depends
on the leptonic kinematics. Nonetheless, in almost all cases, the 
distributions in the case of NP emission are notably different
from the case of PP emission.

\begin{figure*}
	\includegraphics[width=11cm, bb=110 80 520 730]{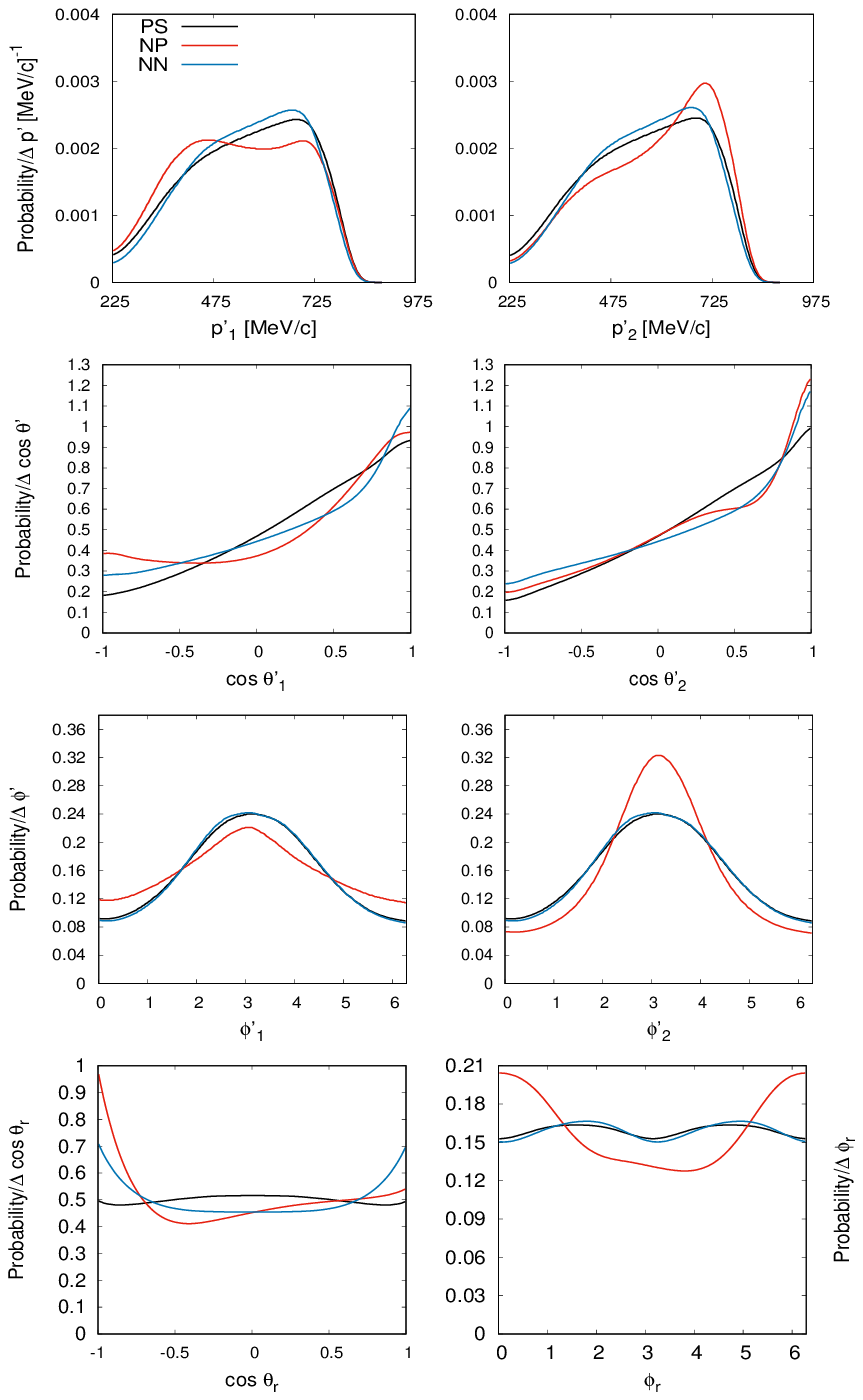}
	\caption{ 
One-fold distributions for semi-inclusive
          two-nucleon emission induced by antineutrinos for kinematics
          1. In each panel, we compare the distributions for
          phase-space (PS), NP, and NN emission.
          }
	\label{sfig8a}
\end{figure*}

\begin{figure*}
	\includegraphics[width=11cm, bb=110 80 520 730]{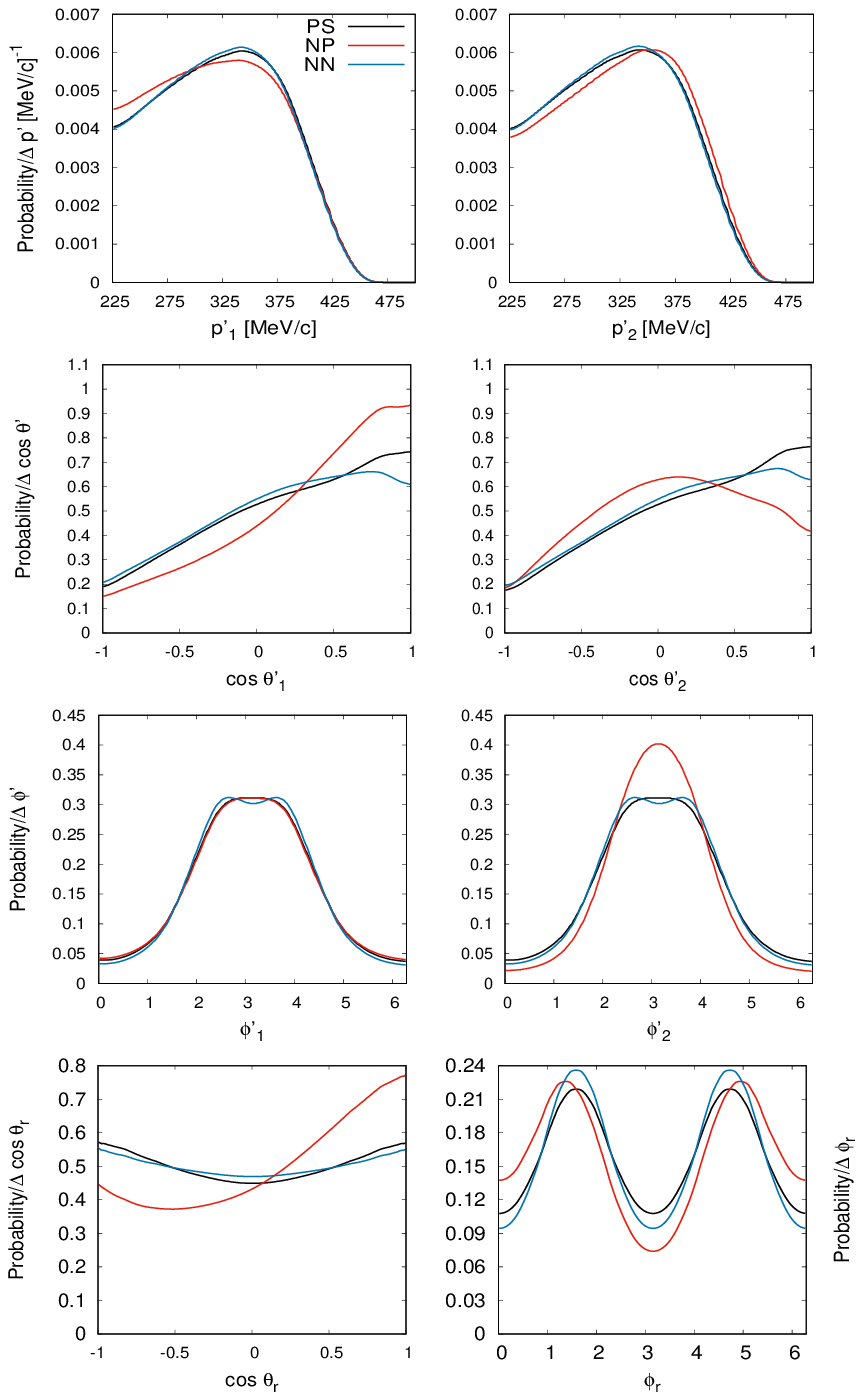}
	\caption{The same as Fig. 12 for kinematics 2.}
	\label{sfig8b}
\end{figure*}

To complete this discussion, we present in Figs. 12 and 13 results for
antineutrino scattering for kinematics 1 and 2, to be compared with
the corresponding figures 10 and 11 for neutrinos. They show similar
trends, with the most noticeable differences occurring again for NP
emission as compared to the NN emission and the phase space
distribution, which are more similar. The most notable differences
between NP and the others are observed in the distributions with
respect to the azimuthal angles.

In any case, it is notable that the significant differences introduced
by the hadronic tensor in the two-fold distributions concerning
relative angles, as shown in Figures 4 and 5, tend to smooth out when
we transition to distributions of the observable variables related to
the momenta of the two nucleons. 
In other words, the very pronounced differences
arising from considering different emission channels compared to the
isotropic distribution in phase space seem to lessen when we consider
the physical variables as a result of an average over bins. It is
expected that these differences become much more pronounced when
considering the complete six-fold distribution, where the full impact
of the hadronic tensor should be more prominent.

\section{Conclusions}

In this work, we have undertaken a comprehensive study of the
semi-inclusive cross section for two-nucleon emission induced by
neutrinos and antineutrinos within a RMF model of
nuclear matter, including MEC. The primary aim of
this study was to assess the differences between various emission
channels, namely NP, PP, and NN, and to compare them with the pure
phase space model commonly employed in neutrino Monte Carlo event
generators.

The novelty in this work lies in the inclusion of the full
semi-inclusive hadronic tensor, derived from a microscopic
calculation, whose integral recovers the inclusive 2p2h hadronic
tensor. Given the complexity of a detailed study of the full
six-dimensional distribution dependent on the final nucleon momenta,
involving numerous variables, we conducted this preliminary
investigation focusing on partial distributions dependent on only one
or two variables (one-fold and two-fold distributions). We integrated
over the remaining variables to gain insights into how the differences
imposed by the hadronic tensor propagate through these observables,
for various lepton kinematics.

Our findings reveal significant variations between the different
emission channels, especially when compared to the pure phase space
model.  These differences are particularly pronounced in distributions
involving the relative angles in the CM frame.  This is evidence that
the isotropic distribution in the center of mass undergoes significant
modifications when the hadronic tensor is included, effectively making
it non-isotropic. This is particularly relevant in the case of NP
emission, which seems to be related to the fact that the NP cross
section is smaller than the PP cross section and is thus more
sensitive to the influence of the hadronic tensor.

In the case of two-fold distributions involving variables of the two
emitted particles, the differences with the pure phase-space model are
smaller due to the averaging effect of integrating over the remaining
variables, which smoothes out the differences. Furthermore, these
distributions are largely determined by kinematics, as energy and
momentum conservation plays a significant role, and the hadronic
tensor mainly influences the finer details of each distribution
without altering the fundamental kinematic constraints.

From our results, we have observed strong correlations between pairs
of variables $(p'_1, p'_2)$, $(\theta'_1, \theta'_2)$, and $(\phi'_1,
\phi'_2)$, which is evident from the corresponding two-fold
distributions. When considering other combinations of two variables,
there is generally very little correlation between them.

Appreciable differences are observed between the NP and PP
distributions when the hadronic tensor is included in neutrino
scattering. One notable result is that when an NP pair is emitted with
neutrinos, it is more likely for the neutron to carry more energy than
the proton, whereas the opposite is true for antineutrino scattering.

We have also presented results for the one-fold distributions, which
were obtained by fixing one semi-inclusive variable and integrating
over the remaining ones. These distributions undergo relatively few
variations when the hadronic tensor is included, primarily due to the
smoothing effect of integration. Nevertheless, significant differences
are observed between the cases of NP and PP emission, for neutrino
scattering, and NP and NN emission, for antineutrino scattering,
depending on the kinematics.

In conclusion, this work has explored semi-inclusive two-nucleon
emission reactions induced by neutrinos, considering a microscopic
treatment of the corresponding hadronic tensor. We have emphasized the
differences with respect to the approach that assumes an isotropic
distribution of the outgoing nucleons in the center of mass. This
research is expected to be valuable for analyses based on Monte Carlo
event generators and offers the potential for improving the
reconstruction of incident neutrino energies.
Future work could extend this study to the full
six-dimensional distribution, providing even deeper insights into the
impact of the hadronic tensor on these processes.

\section*{Acknowledgments}
Work supported by: Grant PID2020-114767GB-I00 funded by
MCIN/AEI/10.13039/501100011033; FEDER/Junta de Andalucia-Consejeria de
Transformacion Economica, Industria, Conocimiento y
Universidades/A-FQM-390-UGR20; and Junta de Andalucia (Grant
No. FQM-225).

\end{document}